\documentclass[12pt]{article}
\usepackage{amsmath}
\usepackage{amssymb}
\usepackage{cancel}
\usepackage{graphicx}

\renewcommand{\thefootnote}{\arabic{footnote}}

\begin{document}
\baselineskip=19pt

\begin{titlepage}

\begin{center}
\vspace*{17mm}

{\Large\bf%
Atmospheric Sterile Neutrinos
}

\vspace*{10mm}
Takehiko~Asaka$^{*\dag}$
~and
~Atsushi~Watanabe$^{\dag}$
\vspace*{10mm}

$^*${\small {\it Department of Physics, 
Niigata University, Niigata 950-2181, Japan}}\\

$^\dag${\small {\it Max-Planck-Institut f\"ur Kernphysik,
Saupfercheckweg 1, 69117 Heidelberg, Germany
}}\\

\vspace*{3mm}

{\small (February 3, 2012)}
\end{center}

\vspace*{7mm}

\begin{abstract}\noindent%
We study production of sterile neutrinos in the atmosphere and their detection
at Super--Kamiokande. 
A sterile neutrino in the mass range $1\,{\rm MeV} \lesssim M_N \lesssim 105\,{\rm MeV}$
is produced by muon or pion decay, and decays to an electron-positron 
pair and an active neutrino. 
Such a decay of the sterile neutrino leaves two electron-like Cherenkov rings 
in the detector.
We estimate the sterile neutrino flux from the well-established active neutrino fluxes
and study the number of the decay events in the detector.
The upper bounds for the active--sterile mixings are obtained by comparing the $2e$-like 
events from the sterile neutrino decays and the observed data by Super-Kamiokande. 
The upper bound for the muon type mixing $\Theta_\mu$ is found to be 
$|\Theta_\mu|^2 \lesssim 5 \times 10^{-5}$ for $20 \,{\rm MeV} \lesssim M_N \lesssim 
80\,{\rm MeV}$, which is significantly loosened compared to the previous estimation. 
We demonstrate that the opening angle and the total energy of the rings
may serve as diagnostic tools to discover the sterile neutrinos in further data 
accumulation and future upgraded facilities.
The directional asymmetry of the events is a sensitive measure of the diminishment of
the sterile neutrino flux due to the decays on the way to the detector.
\end{abstract} 

\end{titlepage}

\newpage
\renewcommand{\thefootnote}{\fnsymbol{footnote}}
\section{Introduction}
\label{intro}
The existence of nonzero neutrino masses has been confirmed in the last
few decades and stimulated the activities aiming for the theory beyond
the Standard Model.
Among various possible ways to introduce the neutrino masses into the model,
adding gauge singlet fermions (right-handed neutrinos) is one of the most economical 
and attractive methods. 
In particular, the seesaw mechanism~\cite{Seesaw} naturally accounts for the observation 
that the neutrino masses are very small compared to the other fermions.
Moreover, it also gives a natural prescription for the baryon asymmetry
of the universe by the leptogenesis scenario~\cite{leptogenesis}.

Besides the usual higher mass scales of the right-handed neutrinos around the 
grand unification scale, the lower mass ranges are also interesting and rich in
phenomenology.
For example, a keV sterile neutrino is a viable dark matter candidate~\cite{DM,
Asaka:2005an} and accounts for the pulsar velocities~\cite{palsar} due to feebleness 
of its interactions.
Two quasi-degenerate sterile neutrinos in the mass range $\mathcal{O}(10^{-1})-
\mathcal{O}(10)\,{\rm GeV}$ provide viable baryogenesis scenario 
alternative to the leptogenesis~\cite{BAU,Asaka:2005pn,BAU2}.
Remarkably, these excellent features originate in a single framework  
so called $\nu$MSM~\cite{Asaka:2005an,Asaka:2005pn}, which is an extension of 
the Standard Model
with three generations of the right-handed neutrinos.
Due to lower threshold energies of the production, 
such ``light'' sterile neutrinos are more likely to be tested in existing and 
forthcoming experiments than the usual right-handed neutrinos with super heavy 
masses (for example, see Ref.~\cite{Gorbunov:2007ak} and references therein).
Mass range by mass range, the sterile neutrinos may provide unique signals in various
circumstances.
It is thus interesting to study how to produce and detect the sterile 
neutrinos.

In this paper, we focus on the sterile neutrinos with the mass range 
$1  \,{\rm MeV} \lesssim M_N \lesssim 105\,{\rm MeV}$ and study their production  
in the atmosphere.
The previous neutrino experiments, including peak searches in the meson decays~\cite{pi,
pi2,pi3,K,K2}
and the decay search with accelerators~\cite{acc,acc2}, have placed certain bounds 
on the active--sterile mixing matrix in this mass range. 
The atmospheric sterile neutrino provides an independent
and complemental test to these experiments with artificial neutrino sources.
In particular, the peak searches of the pion and kaon decays put no stringent bound
on the muon type mixing angle around $M_N = 40\,{\rm MeV}$.
This spot will be effectively probed by the atmospheric sterile neutrino since
the sterile neutrinos in this mass range can be copiously produced by the muon
decays similar to the atmospheric active neutrinos.

The production of sterile neutrinos in the atmosphere and their detection
at Super-Kamiokande (SK) has been discussed in Ref.~\cite{Kusenko}.
In this work, we extend and improve their work in both sterile neutrino production
and its detection.
In the flux estimation, we carefully consider the production of the sterile neutrino
not only by the $\mu^\pm$ decay, but also by the $\pi^\pm$ decay. 
The sterile neutrino flux does not receive the $\pi^\pm$ contributions for 
$M_N \gtrsim 35\,{\rm MeV}$ when the mixing with electron is sufficiently small
(as we will show later, this is indeed the case by
considering the bounds on the mixings from direct search experiments),
being different from the $\nu_\mu $ and $\bar{\nu}_\mu$ fluxes.
We also take into account the energy distribution of the daughter (the sterile neutrino) 
in the parent (mainly muon) decay and the phase space suppression due to the sterile 
neutrino mass.
With the sterile neutrino mass around the neighborhood of the muon mass threshold,
the flux gets suppressed and the mixing bounds are significantly loosened compared 
to Ref.~\cite{Kusenko}.
Furthermore, we study the kinematics of the detection process $N \to e^-e^+ \nu$
in detail, which include the distributions of the opening angle and 
the visible energy of the emitted $e^-$ and $e^+$.
We estimate the upper bounds on the mixing angles by using 1489 days SK data~\cite{
ATMSKI}.
In this estimation, we shall apply the cut to the events by requiring
the visible energy being larger than $30\,{\rm MeV}$ in accordance with the SK data.
This treatment also changes the previous results.

This paper is organized as follows.
In Section~\ref{flux}, we calculate the sterile neutrino fluxes 
produced in the atmosphere.
In Section~\ref{events}, the sterile neutrino decay $N \to e^-e^+ \nu$
and its event rate at SK are studied.
Section~\ref{summary} is devoted to conclusions.

\section{Fluxes of the sterile neutrinos}
\label{flux}
We consider a gauge-singlet fermion $N$ with the mass range 
$1  \,{\rm MeV} \lesssim M_N \lesssim 105\,{\rm MeV}$
which mixes with the left-handed 
neutrino $\nu_\alpha$ ($\alpha = e,\mu,\tau$) as
\begin{eqnarray}
\nu_\alpha \,=\, U_{\alpha i} \,\nu_i + \Theta_{\alpha}N,
\end{eqnarray}
where $U_{\alpha i}$ is the Pontecorvo-Maki-Nakagawa-Sakata (PMNS) matrix,
$\nu_i$ ($i = 1,2,3$) are the mass eigenstates of the active neutrinos.
The parameter $\Theta_{\alpha}$ is the mixing between active and sterile 
neutrinos, which rules the interaction strength of $N$.
The extension to the multi-generation case is trivially done
by replacing $\Theta_{\alpha}N$ with $\sum_I \Theta_{\alpha I}N_I$.
Throughout this work, we focus on the case where the sterile neutrino $N$ is
mainly mixed with $\nu_\mu$ for the shake of simplicity.
That is, we assume $|\Theta_\mu| \gg |\Theta_e|, |\Theta_\tau|$, unless otherwise stated.

Indeed, the electron type mixing $\Theta_e$ is much more severely constrained 
than the other two parameters. 
The peak search with $\pi^\pm \to e^\pm N$ mode suggests $|\Theta_e|^2 \lesssim 
10^{-8} -10^{-7}$ for 
$40\,{\rm MeV}\lesssim M_N \lesssim 140\,{\rm MeV}$, while the decay search
with accelerators indicate the upper bound for $|\Theta_\mu|^2$ varies from
$10^{-4}$ to $10^{-6}$ in the mass rage $10\,{\rm MeV}\lesssim M_N \lesssim 100\,
{\rm MeV}$~\cite{Kusenko}.
In the case of $\Theta_\mu$ dominance, 
the main decay mode of $N$ is $N \to 3\nu$ and the subdominant mode
is $N \to e^-e^+\nu_\mu$ conducted by the neutral currents.
The decay width of each process is given by
\begin{eqnarray}
&&\Gamma(N \to 3\nu) = \frac{G_F^2 M_N^5 |\Theta_\mu|^2}{192\pi^3},
\label{decay1}\\
&&\Gamma(N \to e^-e^+\nu) = \Gamma(N \to 3\nu)\left(
\frac{1}{4} - \sin^2\theta_W +2\sin^4\theta_W \right).
\label{decay2}
\end{eqnarray}
In this paper, we focus on the case where the neutrinos are the Majorana particles.
Then the lifetime of the sterile neutrino is given by $\tau \simeq 
1/2\Gamma(N \to 3\nu)$ $=1.1 \times 10^{-6} \frac{1}{|\Theta_\mu |^{2}}\left( \frac{100 
\,{\rm MeV}}{M_N} 
\right)^5 \,{\rm (s)}$ and the corresponding decay length is $\simeq 0.33 \,\frac{1}
{|\Theta_\mu |^{2}}\left( \frac{100 \,{\rm MeV}}{M_N} \right)^6 \,{\rm (km)}$ for 
$E_N = 100\,{\rm MeV}$ sterile neutrinos.
For $|\Theta_\mu|^2 = 10^{-4}$ and $M_N = 100\,{\rm MeV}$ for example, the decay length 
is about half of the earth radius $R_\oplus \approx 6,400\,{\rm km}$.
The branching ratio of the detectable mode is $\simeq 1/4 - \sin^2\theta_W +2\sin^4
\theta_W = 0.13$.

The sterile neutrinos are produced from charged pions and muons in the same manner
as the active neutrinos.
The main production processes are
\begin{eqnarray}
&&\pi^\pm \to  \mu^\pm \, N,  \nonumber\\
&&\mu^\pm \to e^\pm \nu_e (\bar{\nu}_e) N.  \nonumber
\end{eqnarray}
The former channel is open only for $M_N < m_{\pi^\pm} - m_\mu \approx 35\,{\rm MeV}$.
The kaon contribution is negligible in the energy range of present concern.
In the muon neutrino production for example, $K^\pm$ contribution becomes
significant only in the high-energy regime $E_\nu \gtrsim 100\,{\rm GeV}$~\cite{Lipari}.

In fact, the vital part of the sterile neutrino flux is low-gamma regime
$\gamma \lesssim 10$  for an evaluation of the detectability of the atmospheric 
sterile neutrinos\footnote{Here and henceforth, we use the symbol $\gamma$ to denote 
the gamma factor of the sterile neutrino unless otherwise mentioned.}. 
Namely, we are interested in the sterile neutrino 
spectrum at most up to $1\,{\rm GeV}$ for the mass range of
$1  \,{\rm MeV} \lesssim M_N \lesssim 105\,{\rm MeV}$.
As we will discuss in Section~\ref{events} in detail, the sterile neutrinos are detected
by the decay $N \to e^- e^+ \nu$ which leaves 2$e$-like rings in the detector.
If the gamma factor of the sterile neutrino is too large,
the two fuzzy rings will overlap each other and the separation of the rings
becomes difficult.
Indeed, it turns out that $50$\% ($80$\%) of the whole events will be observed with
the opening angle less then $20^\circ$ for $\gamma = 6\,(10)$.

It is in general a complicated task to calculate the flux for such low-energy
regime since it is affected by solar activity and the geomagnetic field.
In this work, we do not compute the flux from the primary cosmic ray spectrum directly, 
but we try to reconstruct a reasonable parent's (muons and pions) spectrum from 
the well-established active neutrino fluxes available in literature~\cite{Honda,
ATMfluxes}
and then evaluate the sterile neutrino flux from the reconstructed parent fluxes.

The procedure goes as follows.
We start from the $\nu_e$ and $\bar{\nu}_e$ fluxes at SK site~\cite{Honda} to estimate 
the muon fluxes. The evolution of the neutrino flux $\phi_{\bar{\nu}_e}(E_\nu,t)$ is 
described by~\cite{Lipari}
\begin{eqnarray}
\frac{d\phi_{\bar{\nu}_e}(E_\nu,t)}{dt} \,=\,
\int_{E_\nu}^\infty \!\! dE \,\,\frac{\phi_{\mu^-}(E,t)}{\lambda(E,t)}
\,\frac{1}{\Gamma}\frac{d\Gamma}{dE_\nu}(E,E_\nu),
\label{m0}
\end{eqnarray}
where $t$ is the slant depth, $\phi_{\mu^-}(E,t)$ is the muon flux,
$E_\nu\,(E)$ is the neutrino (parent $\mu^-$) energy, 
$\lambda =\rho(t)\sqrt{(E/m_\mu)^2 - 1}/\Gamma$ is the muon decay length multiplied by
the density of the atmosphere $\rho(t)$,
$\Gamma$ is the decay width for $\mu^- \to e^- \bar{\nu}_e \nu_\mu$ in the 
laboratory frame.
By integrating~(\ref{m0}) with respect to $t$, one finds
\begin{gather}
\phi_{\bar{\nu}_e}(E_\nu,t_f) \,=\,
\int_{E_\nu}^\infty \!\!\! dE \,\, \Phi_{\mu^-}(E)\,\frac{1}{\Gamma}
\frac{d\Gamma}{dE_\nu}(E,E_\nu),
\quad\quad
\Phi_{\mu^-}(E) \equiv 
\int_0^{t_f}\!\! dt\frac{\phi_{\mu^-}(E,t)}{\lambda(E,t)},
\label{m1}
\end{gather}
where $t_f$ is the depth corresponding to the sea level.
The sterile neutrino flux $\phi_{N}(E_N,t)$ obeys the similar equation to the active 
neutrinos if the sterile neutrino decay in the atmosphere is negligible.
In fact, the decay length for $N \to 3\nu$ is much larger than the
altitude of the mesosphere edge $\approx 100\,{\rm km}$ for most of the
parameter space. Only exception is the regime $|\Theta_\mu|^2  
\gtrsim 6.6 \times 10^{-3} \left( \frac{\rm 100\,{\rm MeV}}{M_N} \right)^{6}\!
\!\left( \frac{E_N}{\rm 100\,{\rm MeV}} \right)$
where $N \to 3\nu$ is so rapid that some of the sterile neutrinos do not reach 
the detector.
This means for example, the following estimation of the sterile neutrino spectrum
is not reliable and may get additional suppression for $E_N \lesssim 100\,{\rm MeV}$
in the parameter regime $|\Theta_\mu|^2  
\gtrsim 6.6 \times 10^{-3} \left( \frac{\rm 100\,{\rm MeV}}{M_N} \right)^{6}$. 
Putting aside this strong interacting regime, we have 
\begin{eqnarray}
\phi_{N}(E_N,t_f) \,=\,
\int_{E_N}^\infty\!\!dE \, \left( \int_0^{t_f}\! dt\frac{\phi_{\mu^-}(E,t)}
{\lambda'(E,t)}
\right)\, \frac{1}{\Gamma'}
\frac{d\Gamma'}{dE_N}(E,E_N),
\label{phiN}
\end{eqnarray}
where $\lambda'$ and $\Gamma'$ are the decay length and width
for the mode $\mu^- \to e^- \bar{\nu}_e N $.
By using $\lambda' = (\Gamma/\Gamma')\lambda$,
one finds
\begin{gather}
\phi_{N}(E_N,t_f) \,=\,
\frac{\Gamma'}{\Gamma}
\int_{E_N}^\infty\!\!dE \,\, \Phi_{\mu^-}(E)\,\frac{1}{\Gamma'}
\frac{d\Gamma'}{dE_N}(E,E_N),\\
\frac{\Gamma'}{\Gamma} = |\Theta_\mu|^2\left(\,1-8r^2+8r^6-r^8-24r^4\ln(r)\,\right),
\quad\quad r = \frac{M_N}{m_\mu}.
\label{PhaseS}
\end{gather}
Here the integrated muon flux $\Phi_{\mu^-}(E)$ is obtained by fitting the left-hand 
side of~(\ref{m1}) with an assumption of 
power-low behavior of $\Phi_{\mu^-}(E)$. 
The same exercise is applied for $\nu_e$-$\mu^+$-$N$ chain
and then the flux for the Majorana $N$ is obtained as the sum of these two
contributions.
Details for the energy distributions in muon decay are presented in Appendix~\ref{A1}.

In the mass regime $M_N < m_{\pi^\pm} - m_\mu $, pions also contribute to
the sterile neutrino flux.
Roughly speaking, $\pi^\pm$ contribution is the same as
the muon contribution, so that the flux of the sterile neutrino in
$M_N <  m_{\pi^\pm} - m_\mu$ regime is twice as large as the higher mass regime
up to the threshold effect. 
The $\pi^\pm$ contribution is estimated in the same way as the muon.
For the $\pi^\pm$ case, however, we need $\nu_\mu, \bar{\nu}_\mu$ fluxes of
$\pi^\pm$ origin to reconstruct the integrated $\pi^\pm$ fluxes from the muon
neutrino input.
We assess the $\nu_\mu, \bar{\nu}_\mu$ of $\pi^\pm$ origin by subtracting
the $\nu_\mu, \bar{\nu}_\mu$ of $\mu^\pm$ origin from the full 
$\nu_\mu, \bar{\nu}_\mu$ data in Ref.~\cite{Honda}, where $\nu_\mu, \bar{\nu}_\mu$ of 
$\mu^\pm$ origin is calculated from the integrated muon fluxes $\Phi_{\mu^\pm}(E)$
reconstructed via Eq.~(\ref{m1}).
Finally, the sterile neutrino flux of $\pi^\pm$ origin is calculated by
\begin{eqnarray}
\phi_{N}(E_N,t_f) \,=\,
\frac{\Gamma'_\pi}{\Gamma_\pi}
\int_{E_N}^\infty\!\!dE \, \Phi_{\pi^\pm}(E) \,
\frac{1}{\Gamma'_\pi}\frac{d\Gamma'_\pi}{dE_N}(E,E_N),
\end{eqnarray}
where $\Gamma_\pi$ and $\Gamma'_\pi$ are the decay width for
$\pi^\pm \to \mu^\pm \nu_\mu (\bar{\nu}_\mu)$
and $\pi^\pm \to \mu^\pm N$, respectively.
Here the branching ratio is given by
\begin{gather}
\frac{\Gamma'_\pi}{\Gamma_\pi} =
|\Theta_\mu|^2 \sqrt{1 - 2(r_N^2 + r_\mu^2) + (r_N^2 - r_\mu^2)^2}\,
\,\frac{r_N^2 + r_\mu^2 - (r_N^2 - r_\mu^2)^2}{r_\mu^2(1 - r_\mu^2)^2},\nonumber\\
r_N = \frac{M_N}{m_{\pi^\pm}},\quad\quad r_\mu = \frac{m_\mu}{m_{\pi^\pm}}.
\end{gather}
\begin{figure}[t]
\begin{center}
\scalebox{0.15}{\includegraphics{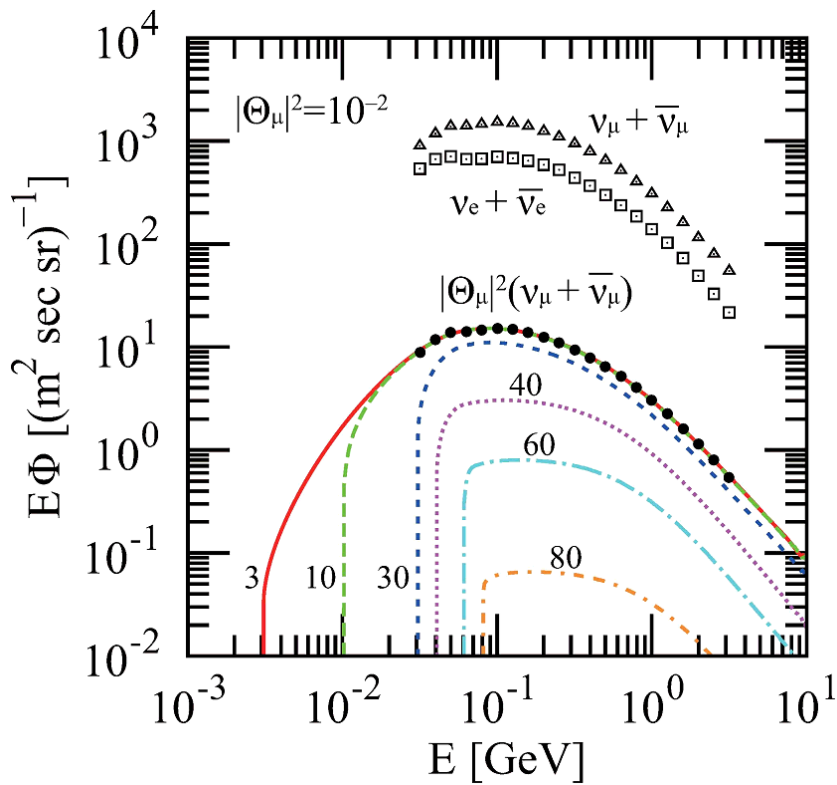}}
\end{center}
\caption{Fluxes of the sterile neutrinos. 
The six curves are for $M_N = 3, \,10, \,30, \,40, \,60, \,80\,{\rm MeV}$ with  
the mixing $|\Theta_{\mu}|^2 = 10^{-2}$.
The triangles and squares show the active neutrino fluxes from Ref.~\cite{Honda}.
The black circles show $|\Theta_\mu|^2\phi_{\nu_\mu + \bar{\nu}_\mu}$.}
\label{fluxA}
\end{figure}
Fig.~\ref{fluxA} shows the sterile neutrino fluxes for several examples of $M_N$.
The six types of the lines are for $M_N = 3, \,10, \,30, \,40, \,60, \,80\,{\rm MeV}$,
respectively (from the left). 
It is seen that a naive estimation $\phi_N = |\Theta_\mu |^2 \phi_{\nu_\mu + 
\bar{\nu}_\mu}$ 
made in Ref.~\cite{Kusenko} is good for the energies above the mass threshold with 
$M_N < 30\,{\rm MeV}$, 
but it breaks down significantly for $M_N \gtrsim 30\,{\rm MeV}$ in such a way that 
the calculated fluxes are suppressed as $M_N$ increases.
This rapid fall off with respect to $M_N$ is due to the phase space suppression 
of the muon decay (See Eq.~(\ref{PhaseS})).
On the other hand, the phase space suppression of $\pi^\pm$ decay is significant only 
at the vicinity of the threshold $m_{\pi^\pm} - m_\mu$. 
As $M_N$ increases, the flux thus suddenly drops by factor of 
two at the threshold $\approx 35\,{\rm MeV}$ above which the $\pi^\pm$ contribution 
vanishes.
The sterile neutrino fluxes presented here are averaged over the zenith angle and
given at the middle level of solar activity following the input neutrino 
fluxes in Ref.~\cite{Honda}.

The uncertainty of the reconstructed parent fluxes $\Phi_{\mu^\pm}$ and $\Phi_{\pi^\pm}$ 
is small for high-energy regime, but it becomes larger as
the energy is lowered so that the resultant sterile neutrino fluxes carry certain
ambiguity around $E_N \simeq M_N$.
This is because high-energy part of the active neutrino flux is well fitted by 
a single-power low, while low-energy part are complex and many choices are available 
as fitting function.
In this analysis, we have used the fitting function 
$\Phi_H \Phi_L/(\Phi_H + \Phi_L)$, where $\Phi_H$ and $\Phi_L$ are the high and
low-energy part of the fitting function.
For $\Phi_H$, we have taken $\Phi_H = a_H \gamma^{b_H}$ with some
constants $a_H,b_H$ and the gamma factor $\gamma$ of the parent particle of interest. 
For $\Phi_L$,  we have examined two options $\Phi_L = a_L \beta^{b_L}$ or 
$\Phi_L = a_L \gamma^{b_L}$, where $\beta$ is the beta factor of the parent particle.  
While with the former $\Phi_L$ the flux rises from the mass threshold and continuously 
shifts to constant behavior and is finally reduced to $\Phi_H$, 
the latter $\Phi_L$ has discontinuity at the mass threshold and the peak is slightly 
shifted to lower energies compared with the former one.
It turns out that two different fitting bring at most $10$\% difference in the total 
number of events.
We would like to emphasize that the salient feature of the phase space 
suppression~(\ref{PhaseS}) is however independent from the fitting scheme and 
the conclusions derived from this effect are valid in what follows.
The fluxes shown in Fig.~\ref{fluxA} are calculated with the option 
$\Phi_L = a_L \beta^{b_L}$ and these fluxes are used in the following analyses.

\section{Decay of the sterile neutrinos and the event rates}
\label{events}
The sterile neutrinos $N$ are produced in the atmosphere and
reach the SK site to leave 2$e$-like events via $N \to e^- e^+ \nu$
and $N \to e^+ e^- \bar{\nu}$.
Suppose that the fiducial volume of SK is represented by a sphere of radius $r$.
The event rate is then given by
\begin{eqnarray}
{\rm Rate} \,= \,
\int_{M_N}^\infty \!\!\!\! dE_N \!\!\int\!\! r^2d\Omega \,\, \phi_N
\,
\int_l^{l + 2r}\!\!\!\!\!dl' \frac{1}{\lambda_d}e^{-\frac{l'}{\Lambda_d}}\,
\int\!\! dX \,\frac{1}{\Gamma_N}\frac{d \Gamma_N}{dX}(E_N, X),
\label{Erate}
\end{eqnarray}
where $\Gamma_N$ and $\lambda_d$ are the decay width and length for the signal decay
of $N$, $X$ is an observable of interest, {\it e.g.,} the invariant mass of the momentums
of $e^-$ and $e^+$, and $l$ is the flight distance of the sterile neutrinos. 
The total decay length $\Lambda_d$ is given by $\Lambda_d = \lambda_d \lambda_d'/
(\lambda_d + \lambda_d')$ where $\lambda_d'$ is the decay length for $N \to 3\nu$.
Since $r$ is much smaller than the earth radius $R_\oplus$, the flight distance $l$ is
well approximated by
\begin{eqnarray}
l \,=\, \left\{ \begin{array}{ccc}
0 & {\rm for} & 0 \leq \theta \leq \frac{\pi}{2},\\
-2R_\oplus\cos\theta   & {\rm for} &  \frac{\pi}{2} < \theta \leq \pi,\\
\end{array}
\right.
\end{eqnarray}
in terms of the zenith angle $\theta$.
The total events are given by the integration over the solid angle 
$\int \!d\Omega = \int_{-1}^1d\!\cos\theta \int_0^{2\pi}d\phi$.
By integrating over all possible final states of $e^-e^+\nu$ and
taking $\Lambda_d \gg 2r$, Eq.~(\ref{Erate}) becomes 
\begin{eqnarray}
{\rm Rate} \,= \,
3V_{\rm fid} \int_{M_N}^\infty \!\!\!\! dE_N \,\frac{1}{\lambda_d}\phi_N
\left[\, \frac{\Lambda_d}{2R_\oplus}\left(\,
1 - e^{-\frac{2R_\oplus}{\Lambda_d}}\,\right) + 1 \,\right],
\end{eqnarray}
where $V_{\rm fid}$ is the fiducial volume of the SK tank $22.5\,{\rm kton}$
which corresponds to $2.25 \times 10^4\,{\rm m^3}$.
Here the first (second) term in the square brackets is the contribution from 
the up (down)-going events. 
For $\Lambda_d \gg 2R_\oplus$, one finds
\begin{eqnarray}
{\rm Rate} \,= \,
6V_{\rm fid} \int_{M_N}^\infty \!\!\!\! dE_N \,\frac{1}{\lambda_d}\phi_N,
\label{Erate2}
\end{eqnarray}
which is independent from the total decay width
and depends only on the partial decay width $\Gamma_N$ responsible
for the detection channel.
The events take place isotropically due to
the condition $\Lambda_d \gg 2R_\oplus$ with which the flight distance $l$
becomes irrelevant to the event number.
While if $\Lambda_d < 2R_\oplus$ so that the sterile neutrinos partially
vanish on the way to the detector, downward-going events dominate over 
upward-going event.
The measurements of the direction of the total momentum of $e^-$ and $e^+$ may
thus clarify the indication of $N \to 3\nu$ decay taking place on the way to the detector.
\begin{figure}[t]
  \centerline{
  \includegraphics[scale=0.107]{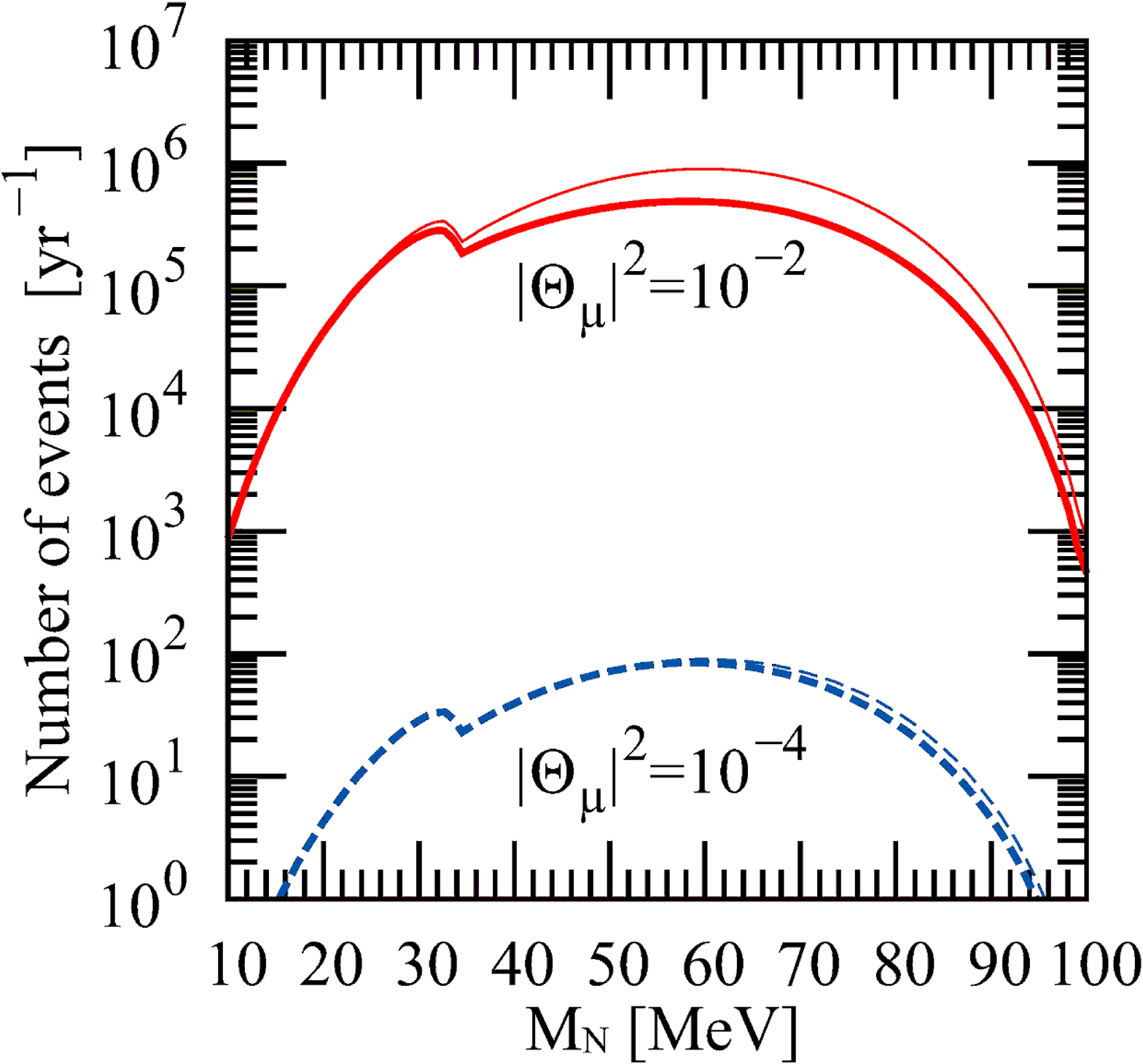}%
  \includegraphics[scale=0.107]{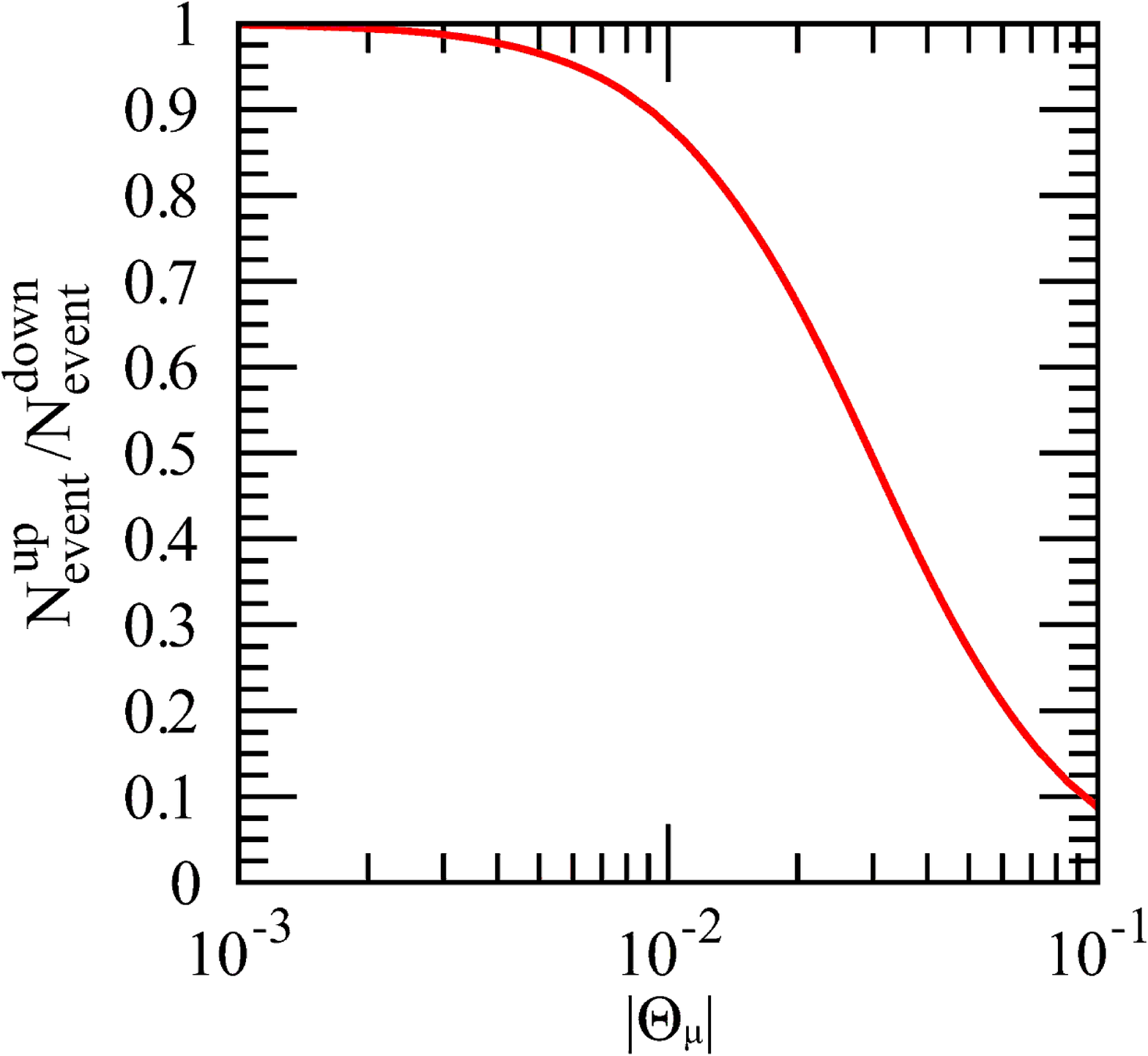}%
  }%
  \caption{Left panel:~Number of events ($N\to e^- e^+ \nu$) at SK per year.
The red solid (blue dotted) lines shows the case when $|\Theta_\mu|^2 = 10^{-2} 
(10^{-4})$. The thick and thin lines are with
and without the effect of $N$ decay from atmosphere to detector.
Right panel:~Up-down asymmetry in number of events. Here we take $M_N = 60$ MeV.
  }
  \label{EventTOT}
\end{figure}

The left panel of Fig.~\ref{EventTOT} shows the total number of events per year 
as a function of the sterile neutrino mass $M_N$.
The red solid (blue dotted) lines shows the case when $|\Theta_\mu|^2 = 10^{-2} 
(10^{-4})$.
It is seen that the event number is proportional to $|\Theta_\mu|^4$.
The thick and thin lines are with and without the effect of $N$ decay from 
atmosphere to the detector.
The thick lines are suppressed compared to the thin lines by the amount
that the up-going events are reduced by the decay effect.
The decay effect is negligible for $|\Theta_\mu|^2 = 10^{-4}$
while it makes certain difference for $|\Theta_\mu|^2 = 10^{-2}$.
The right panel shows the ratio of the up-going events to the down-going events.
As we will see later, the asymmetry is unlikely to be observed for $|\Theta_\mu |^2 
\lesssim 3\times 10^{-5}$ and $M_N = 60\,{\rm MeV}$ where the number of events does 
not exceed the 2$e$-like ring data of SK. 

If $\Theta_\tau$ is switched on, however, it hastens both decay $N \to 3\nu$ and $N \to
e^-e^+\nu$ while the production processes are kept unchanged.
With $\Theta_\tau$ being finite, the decay width of each mode is obtained by 
replacing $|\Theta_\mu|^2$ with $|\Theta_\mu|^2 + |\Theta_\tau|^2$ in~(\ref{decay1}) 
and~(\ref{decay2}).
By virtue of $\Theta_e = 0$, the above two decay modes are induced only by the neutral 
current so that the effect of $\Theta_\tau \neq 0$ simply appears as such a simple 
replacement.
On the other hand, the production processes of $N$ are not affected by $\Theta_\tau$ 
since the decays of $\pi^\pm$ and $\mu^\pm$ are induced by the charged current
and cannot involve a tauon.
Under the assumption that $\Theta_e =0$, the production processes thus involve 
only $\Theta_\mu$.
In the flux equation~(\ref{phiN}), the decay term of $N$ would not be negligible  
if $|\Theta_\tau|$ is large.
This effect however gives negligible contribution to the ratio of upward to 
downward-going events since the thickness of the atmosphere is small compared
to the earth's diameter.

An interesting possibility in view of the up-down asymmetry is therefore
$|\Theta_\tau| \gg |\Theta_\mu|$ with which the flux is feeble while the
decay probability is high. 
In such case the event rate is proportional to $|\Theta_\mu|^2 ||\Theta_\tau|^2$
instead of $|\Theta_\mu|^4$, where $|\Theta_\mu|^2$ comes from the production 
and $|\Theta_\tau|^2$ is from the detection.
Let us set $|\Theta_\mu|^2 ||\Theta_\tau|^2 \sim 10^{-9}$ to obtain 
$\mathcal{O}(10)$ events per year. 
Then, for instance, $|\Theta_\tau |^2 \sim 10^{-2}$ and $|\Theta_\mu |^2 \sim 10^{-7}$
lead to a clear asymmetry $N^{\rm up}/N^{\rm down} \sim 0.1$. (Notice that the right 
panel of Fig.~\ref{EventTOT} can be read as a plot for $|\Theta_\tau|$ in the case where 
$|\Theta_\tau| \gg |\Theta_\mu|$.)
If the asymmetry is observed, it may indicate a hierarchical structure of
$|\Theta_\tau| \gg |\Theta_\mu|$.

The master formula for the event rate~(\ref{Erate}) involves the product of 
the production and the detection probabilities.
That is, the product of the squared modulus of the production and the detection 
amplitudes.
This means that~(\ref{Erate}) is valid in the case where the coherence between the 
propagating neutrino states is lost.
Since we have the sterile neutrino and the active neutrinos in the theory,
one may wonder if the oscillation between the active and sterile states occurs.
However, the coherence between the active and sterile states is explicitly violated
at the detection point; the sterile neutrino decays while the active neutrinos do not.
Furthermore, even if the coherence were not violated by the detection, 
such an oscillation would be so rapid that the effect is averaged out, which also
insures the validity of using~(\ref{Erate}).
In fact, with $M_N = 10\,{\rm MeV}$ and a typical energy of $100\,{\rm MeV}$ for example, 
the oscillation length $L^{\rm ocs}$ would be
$L^{\rm ocs} \simeq \left( \frac{E}{100 \,{\rm MeV}} \right)
\left( \frac{{100\,\rm MeV^2} }{ \Delta m^2} \right)10^{-12} \,\,{\rm m}$.
In the present setup, it is quite difficult to observe oscillatory behavior between
the active and the sterile neutrinos.

\subsection{Invariant mass and  bounds for the mixing}
\begin{figure}[t]
\begin{center}
\scalebox{0.13}{\includegraphics{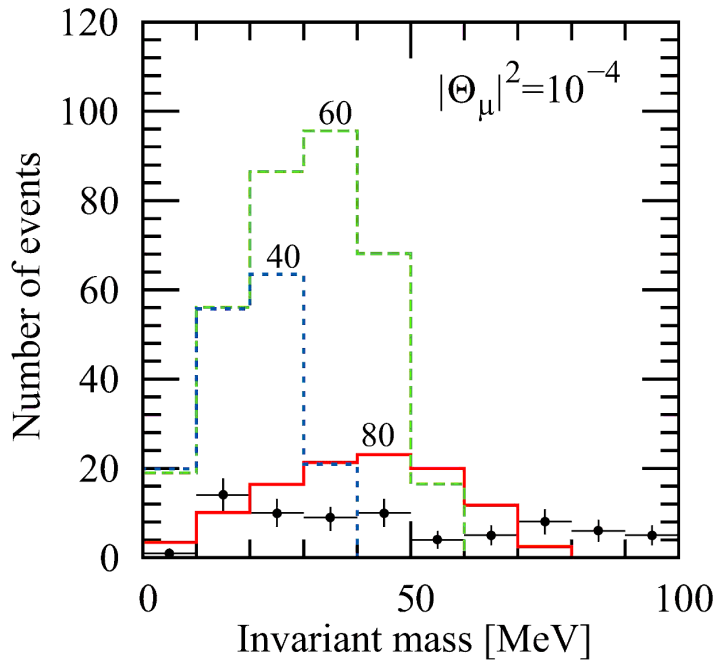}}
\end{center}
\caption{Invariant mass distribution of the electron pairs from
the sterile neutrino decay $N \to e^- e^+ \nu$.
The solid (red), dashed (green), and short-dashed (blue) lines
are for $M_N = 80,\,60,\,40\,{\rm MeV}$ with the mixing $|\Theta_\mu|^2 =
10^{-4}$.
The points with error bars are the data of fully-contained
$2e$-like rings from Ref.~\cite{ATMSKI}.
}
\label{EventMee}
\end{figure}
The sterile neutrino decay $N \to e^-e^+\nu$ in the SK detector produces two
fuzzy Cherenkov rings.
A possible background for this signal is $\pi^0 \to 2\gamma$, where $\pi^0$ are
mainly created via the neutral current interactions of the atmospheric active neutrinos.
The gamma rays in the final state develop into electromagnetic showers and create
two fuzzy rings which cannot be distinguished from the $e^\pm$ signals.

The above two process, however, leave different signatures in the Lorentz invariant 
mass (squared) of the $2e$-like rings.
For the sterile neutrino event $N \to e^-e^+\nu$, it is given by $M_{ee}^2 = 
(p_1 + p_2)^2$, where $p_1(p_2)$ are the four-momenta of $e^-(e^+)$.
In the rest frame of the sterile neutrino, it reads
\begin{eqnarray}
M_{ee}^2 \,\simeq\, M_N^2 -2M_N E_{\nu},
\end{eqnarray}
where $E_\nu$ is the invisible energy carried away by the active neutrino.  
Since the neutrino carries about $1/3$ of the parent energy on average, 
$M_{ee}^2$ from the sterile neutrino events follows $\langle M_{ee}^2 \rangle 
\simeq M_N^2/3$, while the invariant mass squared of the two photons by $\pi^0$ decay
sharply peaks at $m_{\pi^0}^2$.  
By measuring the invariant mass distribution, we may obtain information 
on the sterile neutrino mass.

It is more practical to work with $M_{ee}$ than $M_{ee}^2$.
The $M_{ee}$ distribution is given by (See Appendix~\ref{A3} for details)
\begin{eqnarray}
\frac{1}{\Gamma_N}\frac{d\Gamma_N}{d M_{ee}}\,=\,
\frac{4}{M_N}y_{ee}(1-y_{ee}^2)^2(1+2y_{ee}^2),
\label{MeeDis}
\end{eqnarray}
where $y_{ee} = M_{ee}/M_N$.
The distribution~(\ref{MeeDis}) has the maximum at $y_{ee} \sim 0.5$,
and the averaged invariant mass is given by
$\langle M_{ee} \rangle = \int_0^{M_N} dM_{ee} \frac{M_{ee}}{\Gamma_N}\frac{d\Gamma_N}
{dM_{ee}} = 0.508\,M_N$.
Note that the distribution function~(\ref{MeeDis}) also holds in the laboratory 
frame due to the Lorentz invariance of $M_{ee}^2$.
We calculate the event rate for each invariant mass by using~(\ref{MeeDis})
in~(\ref{Erate}).

Fig.~\ref{EventMee} shows the number of events in each invariant mass bin.
The solid (red), dashed (green), and short-dashed (blue) lines
are for $M_N = 80,\,60,\,40\,{\rm MeV}$ with the mixing $|\Theta_\mu|^2 = 10^{-4}$.
The points with error bars are the $1489$ days data of fully-contained
$2e$-like rings from Ref.~\cite{ATMSKI}.
Following~(\ref{MeeDis}), the events are distributed to the range $0<M_{ee}<M_N$ 
and frequently seen at $\simeq 0.5 M_N$.
From $M_N = 40\,{\rm MeV}$ to $M_N = 60\,{\rm MeV}$, the hight of the peak grows
although the possible range of $M_{ee}$, to which the events are distributed, 
is broadened.
This is because the total number of event increases from $M_N = 40\,{\rm MeV}$ to 
$M_N = 60\,{\rm MeV}$ (see Fig.~\ref{EventTOT}).
On the other hand, the hight of the peak rapidly falls off from $M_N = 60\,{\rm MeV}$
to $M_N = 80\,{\rm MeV}$ since  not only the range of $M_{ee}$ is broadened but also 
the total number of event decreases.

\vspace{3mm}
\begin{figure}[t]
  \centerline{
  \includegraphics[scale=0.107]{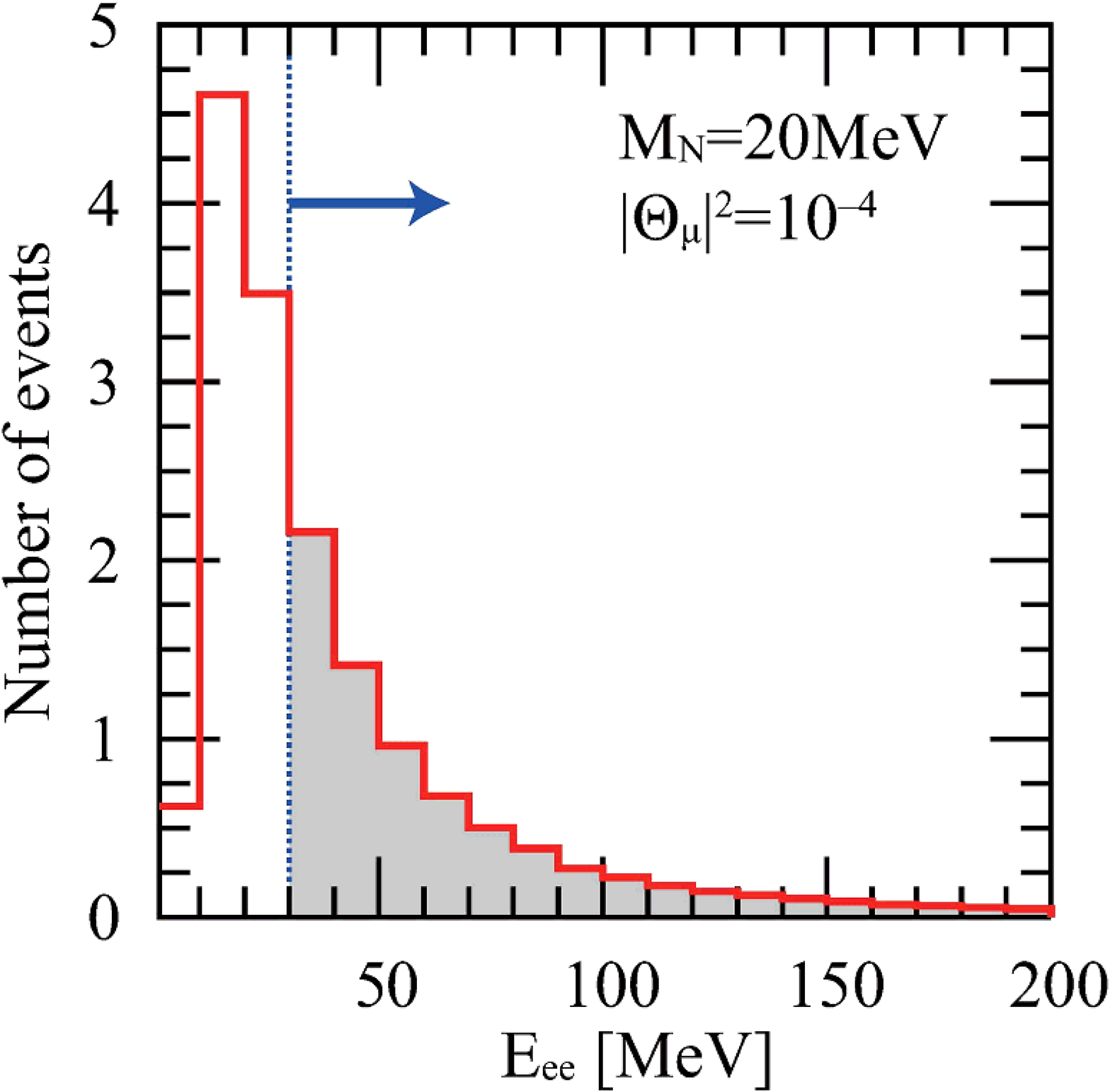}%
  \includegraphics[scale=0.107]{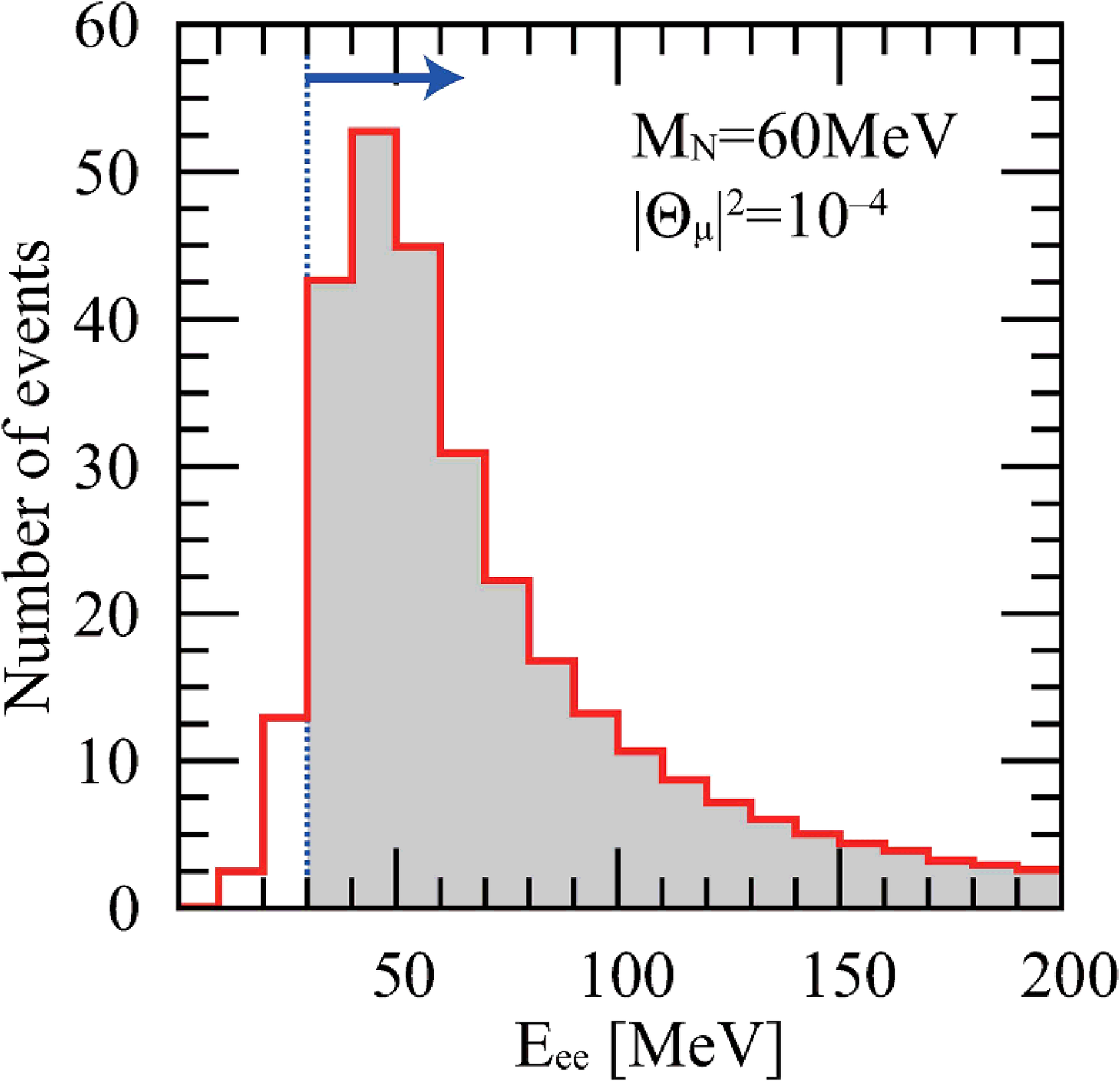}%
  }%
  \caption{Visible energy distributions for $M_N = 20\,{\rm MeV}$ (left)
and $M_N = 60\,{\rm MeV}$ (right). Here we take $|\Theta_\mu|^2=10^{-4}$.
}
  \label{EvisFig}
\end{figure}

A naive upper bound for $\Theta_\mu$ would be determined by comparing the hight
of each bin with the corresponding background for fixed values of $M_N$.
However, it should be noticed that the SK data shown in Fig.~\ref{EventMee} is 
for visible energy greater than $30\,{\rm MeV}$, where the visible energy
means the total energy of the two rings (hereafter we denote it by $E_{ee}$). 
Hence we must cut the events whose $E_{ee}$ is less than $30\,{\rm MeV}$ to make 
such a comparison.
In addition, 
In addition, the opening angle between two momenta of $e^-$ and $e^+$ must be
sufficiently large in order for the event to be identified as multi-ring events.
Thus one must study how $E_{ee}$ and the angle cut reduce the number of
events for a reasonable estimation of the allowed parameter range.

Roughly speaking, the two cut define an effective range of the sterile neutrino 
energy $E_N$ for a given value of $M_N$. 
In particular, the visible-energy cut defines a lower limits of $E_N$.
In the three-body decay $N \to e^-e^+\nu$, the electron and positron carry
on average about $1/3$ of the parent energy each, so that the events with 
$E_N \gtrsim 45\,{\rm MeV}$ more or less pass the cut.
Accordingly, the number of event receives significant reduction for $M_N \lesssim
 45\,{\rm MeV}$ while it does not for $M_N \gtrsim 45\,{\rm MeV}$.
Fig.~\ref{EvisFig} shows two examples of $E_{ee}$ distributions. 
It is seen that about $60$\% of the whole event is dropped for $M_N = 20\,{\rm MeV}$ 
but the cut is insignificant for $M_N = 60\,{\rm MeV}$.

\begin{figure}[t]
  \centerline{
  \includegraphics[scale=0.107]{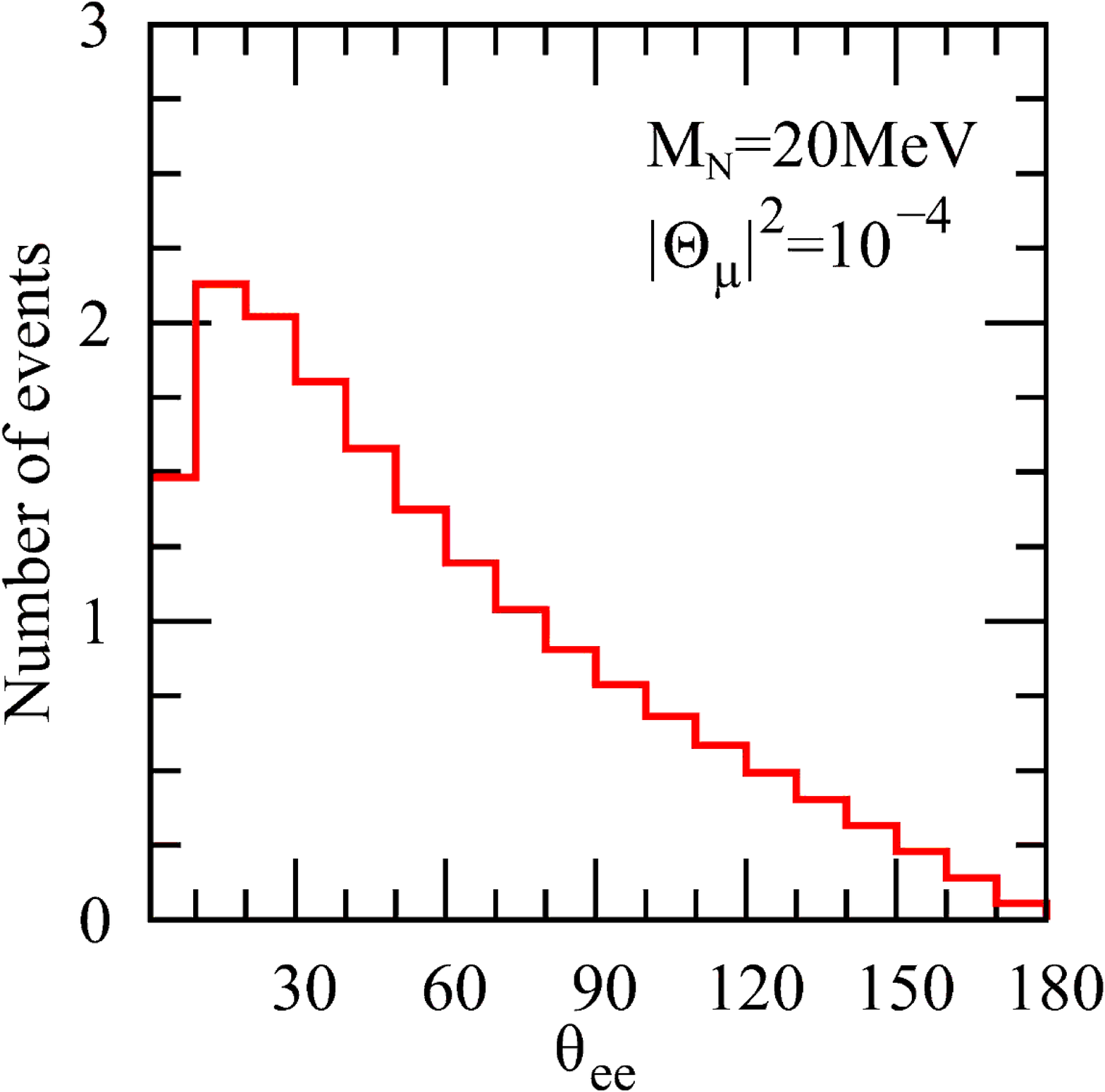}%
  \includegraphics[scale=0.107]{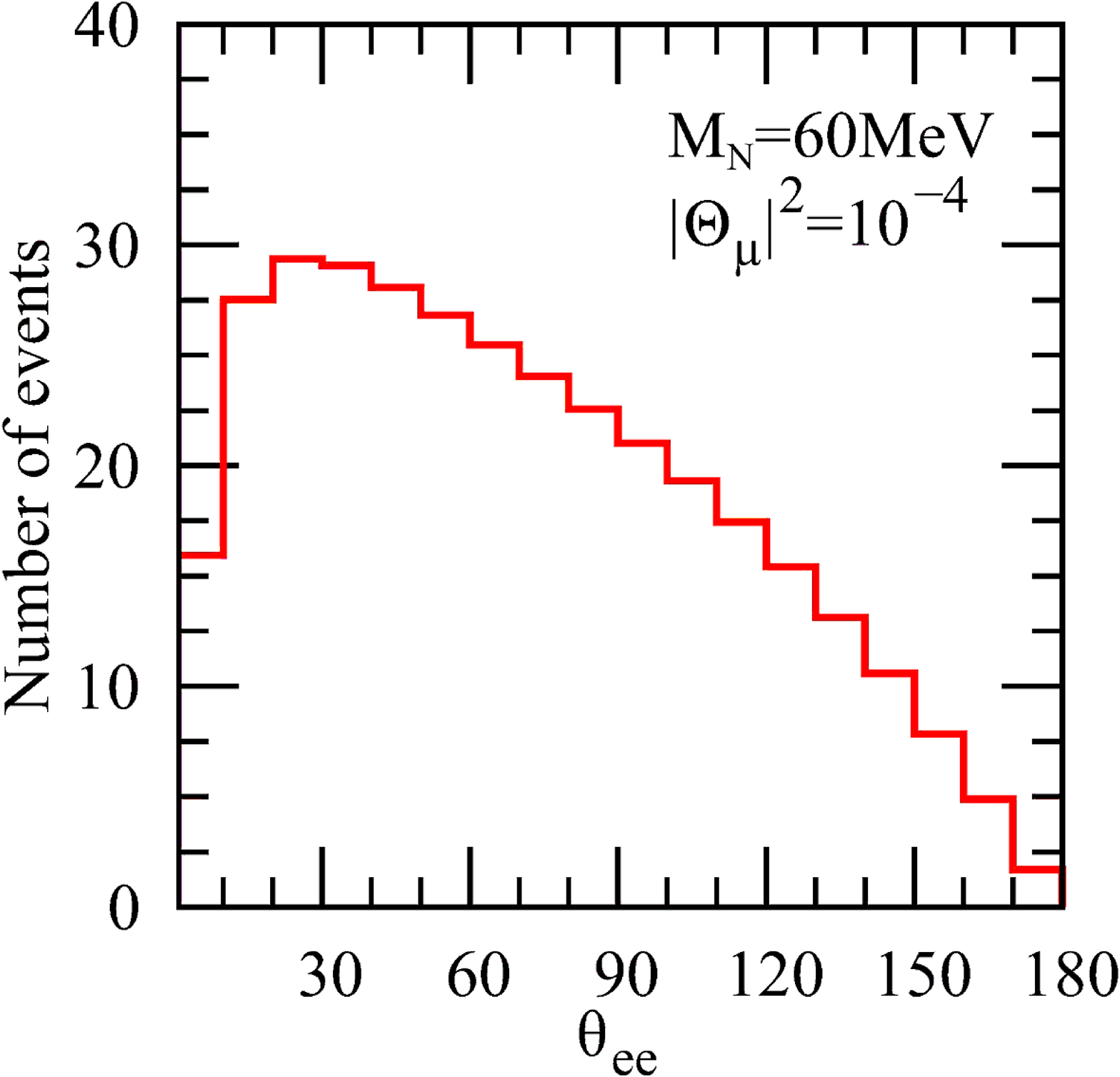}%
  }%
  \caption{
Opening angle $\theta_{ee}$ distributions for $M_N = 20\,{\rm MeV}$ (left)
and $M_N = 60\,{\rm MeV}$ (right). Here we take $|\Theta_\mu|^2=10^{-4}$.
}
  \label{AngleFig}
\end{figure}

On the other hand, the angle cut does not dramatically change the number of events.
In Fig.~\ref{AngleFig}, we show the opening-angle distributions for $M_N = 20\,{\rm MeV}$
and $M_N = 60\,{\rm MeV}$, assuming angle resolution of $10^\circ$.
It is clear that the angle cut does not have as much impact as the visible energy cut
for both cases if the two rings with $\theta_{ee} \gtrsim 10^\circ$,
or more conservatively, $\theta_{ee} \gtrsim 20^\circ$, can be separately identified
at SK.
We found that efficiencies of the angle cut are $0.65$ for $M_N = 10\,{\rm MeV}$,
$0.8$ for $M_N = 20\,{\rm MeV}$ and $0.85$ for $M_N = 30 - 100\,
{\rm MeV}$ with $\theta_{ee} > 20^\circ$. 
The sterile neutrinos with smaller mass receives more reduction than 
larger mass since light sterile neutrinos are more energetic.
It is seen in Fig.~\ref{fluxA} that sterile neutrinos are mostly populated
in $\gamma \sim 4-5$ for $M_N = 20\,{\rm MeV}$ while $\gamma \sim 1- 2$
for $M_N = 60\,{\rm MeV}$. 

\begin{figure}[t]
  \centerline{
  \includegraphics[scale=0.12]{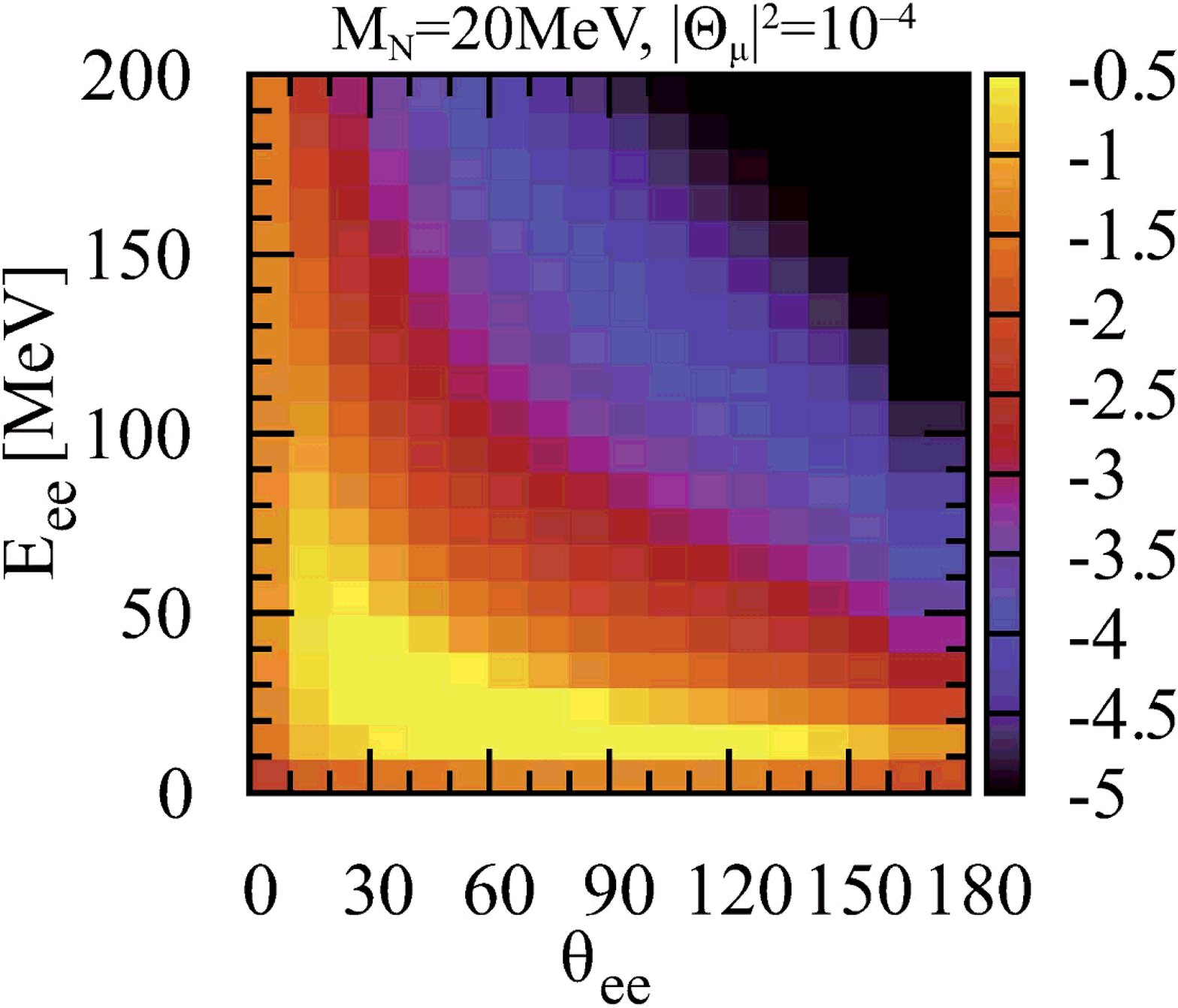}%
  \includegraphics[scale=0.12]{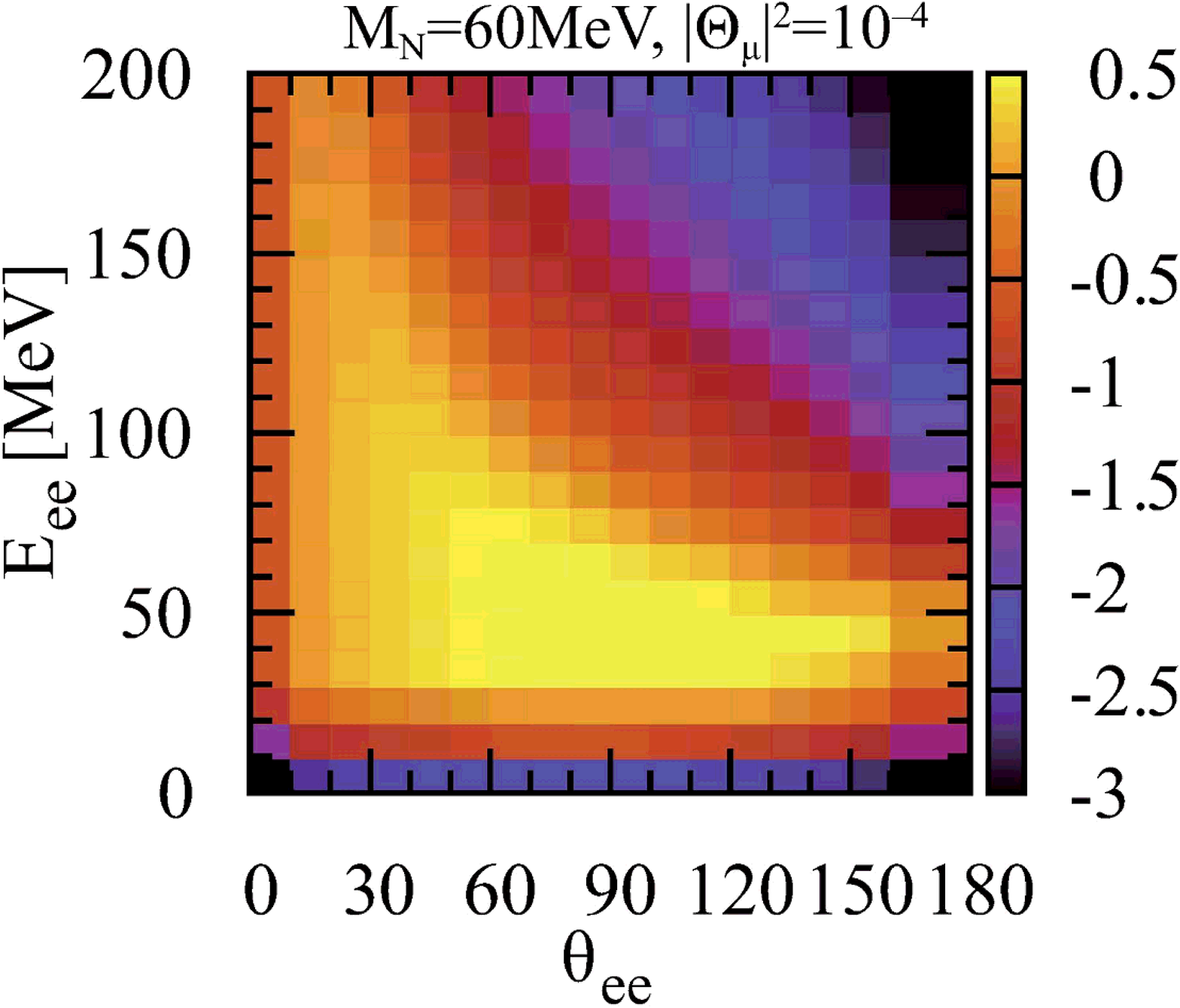}%
  }%
  \caption{
Correlation between $E_{ee}$ and $\theta_{ee}$ for $M_N = 20\,{\rm MeV}$ (left)
and $M_N = 60\,{\rm MeV}$ (right). The colors in each bin shows $\log_{10}(
{\rm number\, of\, events})$.
$|\Theta_\mu|^2=10^{-4}$ is taken.
}
  \label{EvisAngleFig}
\end{figure}

We present correlations between $E_{ee}$ and $\theta_{ee}$ in Fig.~\ref{EvisAngleFig}
for completeness.
For a lighter mass $M_N = 20\,{\rm MeV}$, most sterile neutrinos are energetic $\gamma 
\sim 4-5$ and the events are concentrated on lower-left region.
For the heavier mass $M_N = 60\,{\rm MeV}$, the peak is shifted toward larger 
$\theta_{ee}$ region since there are many sterile neutrinos with $\gamma \sim 1$,
and $\theta_{ee}$ tends to be close to $\sim 120^\circ$ expected from the decay at rest.
We would like to emphasize that these plots may be useful to separate the signal
from the background if the tendency of the background is different from the signal.

\begin{figure}[t]
  \centerline{
  \includegraphics[scale=0.107]{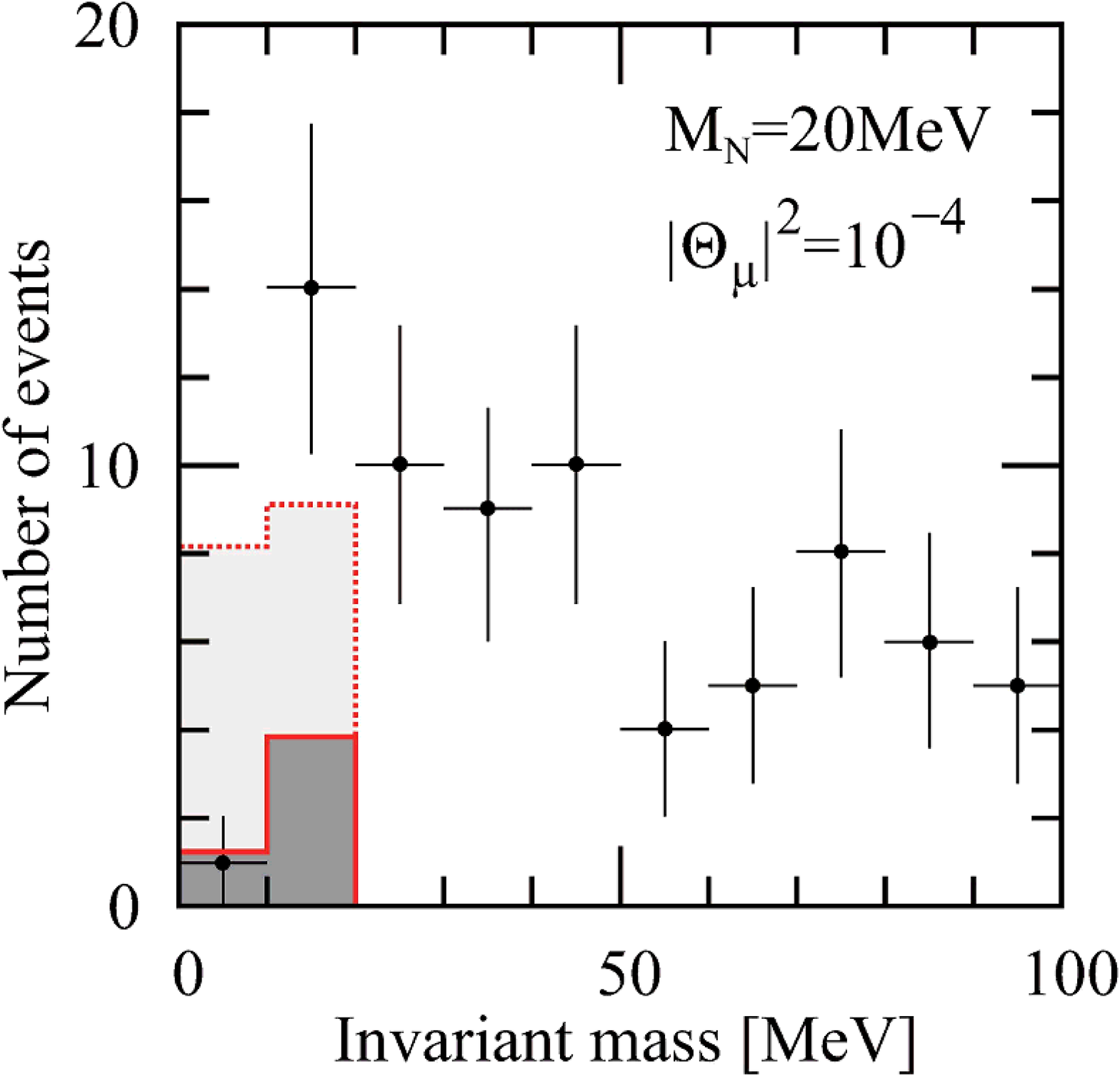}%
  \includegraphics[scale=0.107]{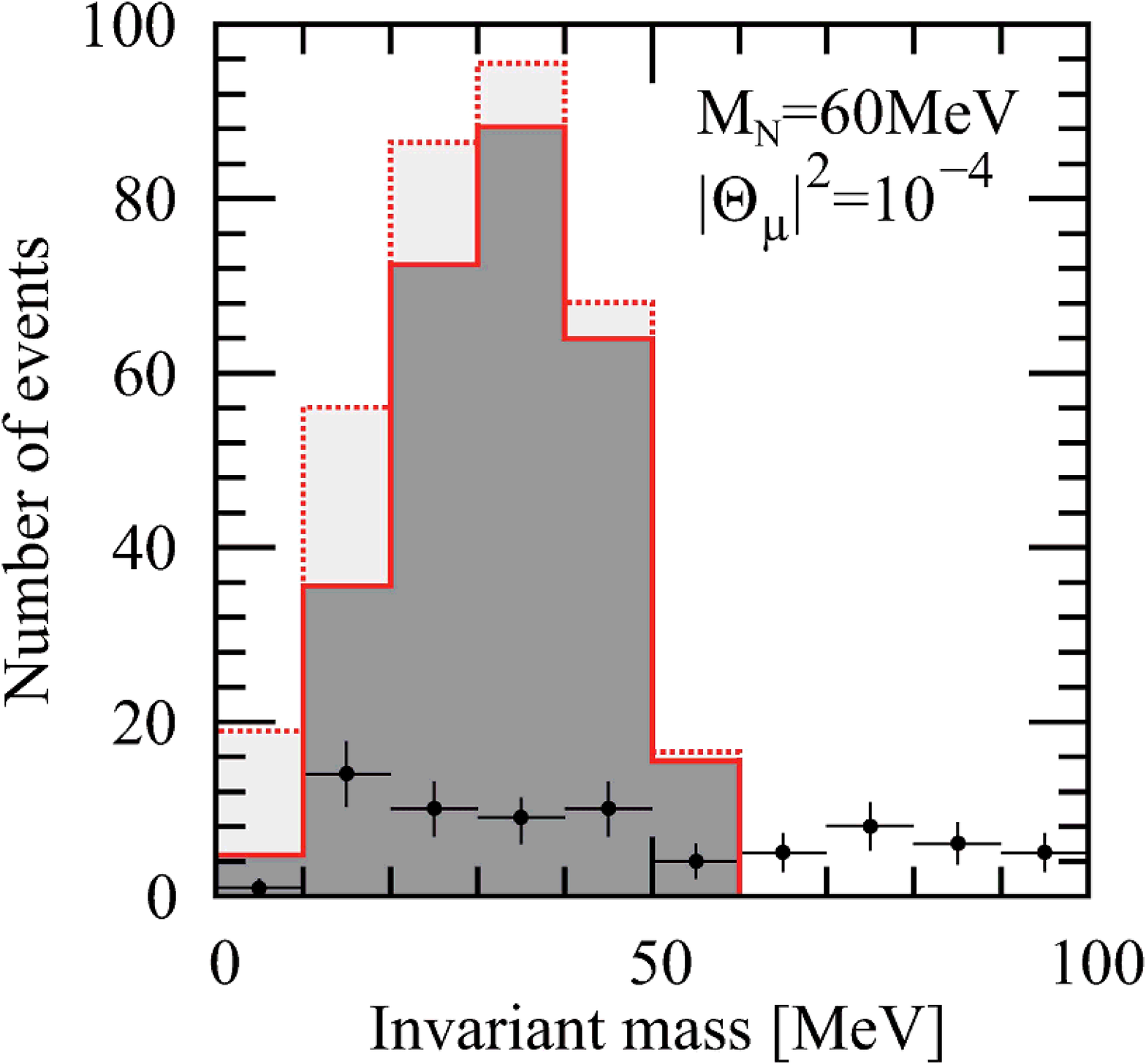}%
  }%
  \caption{
Invariant mass distributions of envents at SK for
    $M_N = 20$ MeV (left) and $M_N=60$ MeV (right).
    The light gray region with red dotted line
    is without cuts, while 
    the dark gray region with red solid line
    is with $\theta_{ee} \ge 20^\circ$ and $E_{ee}\ge 30$ MeV.}
\label{MeeCutFig}
\end{figure}
\begin{figure}[t]
\begin{center}
\scalebox{0.15}{\includegraphics{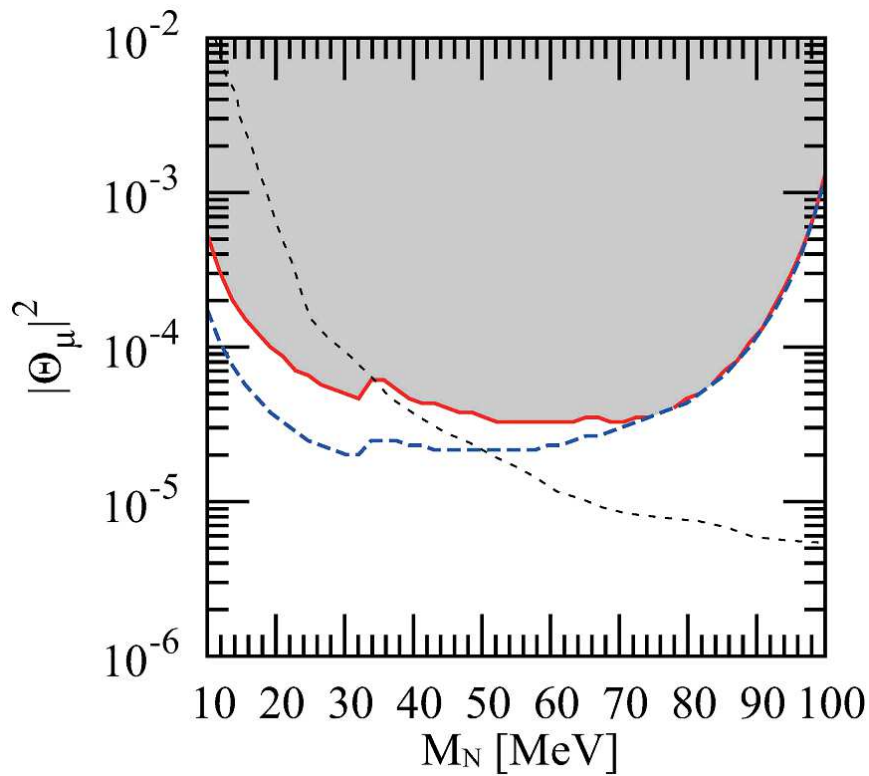}}
\end{center}
\caption{
Upper bounds on the mixing element $|\Theta_\mu|^2$.  The
    red solid and blue dashed lines are bounds with and without cuts
    $\theta_{ee} \ge 20^\circ$ and $E_{ee}\ge 30$ MeV, respectively.
    The black dotted line shows the bound in Ref.~\cite{Kusenko}.}
\label{BoundFig}
\end{figure}
\begin{figure}[t]
\begin{center}
\scalebox{0.15}{\includegraphics{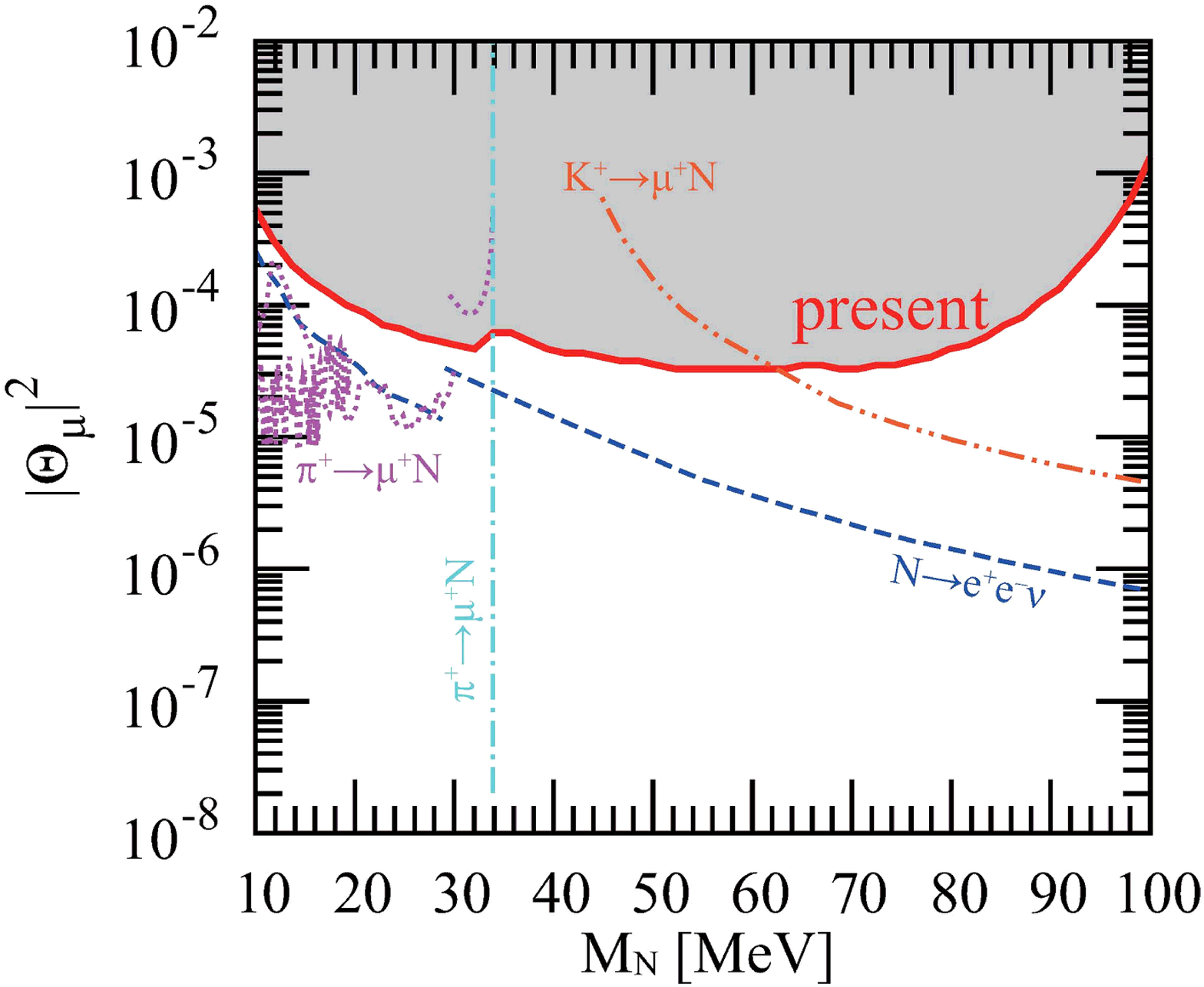}}
\end{center}
\caption{
Upper bounds on the mixing element $|\Theta_\mu|^2$ (the red solid curve) in 
comparison with vrious experiments. 
The dotted (purple) curve is from the peak search in pion decay~\cite{pi2} (labeled by 
$\pi^+ \to \mu^+ N$), the dot-dashed (light-blue) line is from the peak search in pion 
decay~\cite{pi3}, the dashed curve (blue) is from the accelerator decay search by 
PS191~\cite{acc2} (labeled by $N \to e^+e^-\nu$), and the dot-dot-dashed (orange) curve 
is from the peak search in kaon decay~\cite{K2} (labeled by $K^+ \to \mu^+ N$).}
\label{BoundFig2}
\end{figure}

\vspace{3mm}

Having details for the visible energy and the opening angle distribution,
let us come back to the invariant mass and estimate the upper bound of the mixing 
element $\Theta_{\mu}$.
Fig.~\ref{MeeCutFig} shows the invariant mass distributions of events 
with a kinematical cut.
As is already mentioned, lighter-mass case receives more significant reduction
by the cuts. 
We would like to stress that, while the visible energy cut $E_{ee} \geq 30\,{\rm MeV}$ 
is indeed applied to the SK data, the angle cut is not the actual one which is applied to
the data. 
In fact, the multi-ring identification process performed in the 
SK analysis~\cite{Ishihara} is much more complicated than the selection of opening 
angles  and such a thorough analysis is beyond the scope of this paper.
Since we do not know the minimal opening angle appropriate for the SK detector
and could not find suitable references for this value, here and henceforth we assume 
it to be $20^\circ$ just for example.
We estimate the upper bound for $\Theta_{\mu}$ by searching for the maximal value of 
$\Theta_{\mu}$ with which none of the signal events of each bin exceeds the central 
value of the data~\cite{ATMSKI} (shown by dots with error bars in the figures) 
for each fixed value of $M_N$.

Fig.~\ref{BoundFig} shows the upper bound in comparison with the previous 
result in Ref.~\cite{Kusenko}. 
The bound is changed in three ways.
First of all, in the large-mass region $M_N \gtrsim 50\,{\rm MeV}$, the bound is 
significantly relaxed due to the phase space suppression of the muon decay which has 
not been taken into account in the previous analysis.
Second, in the small-mass region $M_N \lesssim 50\,{\rm MeV}$, the event cuts (mainly
the visible energy cut $E_{ee} > 30\,{\rm MeV}$) reduce the number of events so that
the bound is pushed up from the curve without the cuts. 
Finally, careful handling of the $\pi^\pm$ and $\mu^\pm$ contributions for
the sterile neutrino production makes a small dip at $M_N = m_{\pi^\pm} - m_\mu$ 
since the number of events are reduced for $M_N > m_{\pi^\pm} - m_\mu$ according to 
the absence of the $\pi^\pm$ contributions.

The light-element abundances predicted by the Big Bang Nucleosynthesis (BBN) is kept
unspoiled if the lifetime of $N$ is short enough such that the
sterile neutrinos are cleared away before the onset of the BBN. 
According to Ref.~\cite{BBN}, the successful BBN requires
\begin{eqnarray}
|\Theta_\alpha|^2 \,>\, 568.4 \left( \frac{M_N}{\rm MeV} \right)^{-3.549}
\!\!\!\!\!\! - 5.17\times
10^{-6},  \quad\quad(\alpha = \mu, \tau),
\label{BBNlim}
\end{eqnarray}
for $10\,{\rm MeV} < M_N < 140\,{\rm MeV}$.
Here this limit is valid for the Dirac type sterile neutrino which mixes 
with only one active flavor.
For the Majorana case, this limit 
becomes two times weaker since the total decay width of the Majorana particle involves 
an extra factor of two.
Most of the allowed region in Fig.~\ref{BoundFig} is in fact under tension with the 
limit~(\ref{BBNlim}).
However, this tension is relaxed if $\Theta_\tau$ is turned on.
Namely, the lifetime is shortened by increasing $|\Theta_\tau|$.
The point is that the production of the sterile neutrino in the atmosphere is not
affected by $|\Theta_\tau|$.
Hence, by taking $|\Theta_\tau| \gg |\Theta_\mu|$, one can set a  short lifetime enough
to avoid the BBN limit while keeping the atmospheric sterile neutrinos consistent with
the SK data.
Notice that  Fig.~\ref{BoundFig} can be read as a plot for $|\Theta_\mu||\Theta_\tau|$ 
if $|\Theta_\tau| \gg |\Theta_\mu|$.
Interestingly, the directional asymmetry (see Fig.~\ref{EventTOT} and related 
discussion) is likely to be observed in such a circumstance.

If the visible energy cut $E_N > 30\,{\rm MeV}$ is switched off, the sterile neutrino 
events in the first three bins increase as shown in Fig.~\ref{MeeCutFig}, while the 
background and its Mote Carlo simulation might change differently depending on the
source of the background.
Such an analysis is important especially for light sterile neutrinos with $M_N < 30\,
{\rm MeV}$ but beyond the scope of this paper.

\section{Conclusions}
\label{summary}
We have studied the production and the detection of the atmospheric sterile neutrinos.
With the mass range $1\,{\rm MeV} \lesssim M_N \lesssim 105\,{\rm MeV}$,  
the sterile neutrinos are produced by muon or pion decays. 
The main decay mode of the sterile neutrino $N$ is $N \to 3\nu$ ($88$\%)
and the subdominant mode is $N \to e^-e^+\nu$ ($12$\%).
Interestingly, the detection via the subdominant channel is feasible 
at Super-Kamiokande.
To estimate the amount of the sterile neutrino flux $\phi_N$, we have reconstructed 
the parent muon and pion integrated fluxes from the well-established active neutrino 
fluxes $\phi_\nu$ by making a power-low ansatz for the shape of the spectrum.
We have calculated the sterile neutrino flux from the reconstructed parent fluxes
and the energy distributions of each decay mode.
The phase space suppression of the muon decay $\mu^\pm \to e^\pm \nu_e (\bar{\nu}_e)N$ 
is so effective that the estimated sterile neutrino flux is much less than the
naive expectation of $\phi_N = |\Theta_\mu |^2 \phi_{\nu + \bar{\nu}_\mu}$ for $M_N 
\gtrsim 40\,{\rm MeV}$.

For the detection, we have calculated the number of the decay event $N \to e^-e^+\nu$ 
at SK.
The upper bound for the muon-type mixing $\Theta_\mu$ is estimated by comparing 
the invariant mass distribution of the $e^\pm$ pair and the observational 
data of the 2$e$-like rings.
To estimate the bound, the visible energy cut $E_N > 30\,{\rm MeV}$ is taken into
account in accordance with the SK data.
Moreover, we have studied the opening angle of the $e^\pm$ momenta and clarify the 
impact of the minimal opening angle under which the two fuzzy rings cannot be
identified separately.
The upper bound for higher mass regime $M_N \gtrsim 50\,{\rm MeV}$ is significantly 
relaxed compared to the previous estimation due to the phase space suppression of
the muon decay.
In addtion, the inclusion of the kinematical cuts also relaxes the bound
for lower mass regime and changes the shape of the bound entirely.

Fig.~\ref{BoundFig2} shows the upper bound in comparison with the other 
experiments (see for example, Ref.~\cite{Nsearch1,Ruchayskiy:2011aa}).
In the mass window of $34\,{\rm MeV} < M_N \lesssim 64 \,{\rm MeV}$, the atmospheric 
bound is stronger than the peak search in the meson decays. 
However, the bound from the accelerator decay search\footnote{~We have included the 
neutral current contribution to $N \to e^-e^+\nu$ to apply the bound from 
Ref.~\cite{acc2}.
See Ref.~\cite{Kusenko,Ruchayskiy:2011aa}.} is stronger than the
atmospheric bound by a factor of two at $M_N = 34\,{\rm MeV}$ and by a factor of 10 at 
$M_N = 64\,{\rm MeV}$.
We nevertheless believe that the search for the atmospheric sterile neutrino is an issue 
of great interest. 
One reason is that not a few progresses in particle physics have been made 
in cooperation between artifical and natural source experiments.
Since PS191 is the sole experiment which set the bound stronger than the 
atmospheric one in the window $34\,{\rm MeV} < M_N \lesssim 64 \,{\rm MeV}$,
the atmospheric sterile neutrino can play a role of a unique follow-up experiment with 
a natural source. 

Moreover, many points are to be ameliorated for better sensitivities of the atmospheric 
sterile neutrino.
It is obvious that our naive scheme to estimate the upper bound does not reflect the 
accumulation of statistics and more advanced analysis should be performued
for precise argument.
The 2$e$-like ring data for SKI, SKI\!I and SKI\!I\!I~\cite{Ishihara} can be combined
to squeeze the room remained for the sterile neutrino.
Furthermore, the analysis of the 2$e$-like events can be customized aiming for
the sterile neutrino detection.
For instance, one can switch off the visible energy cut of $30\,{\rm MeV}$ 
and track the direction of the sum of the two ring's momenta, the opening angle
between two rings and the visible energy.
This may dramatically change the sensitivity if the background behaves differently
from the signal in the up/down-going ratio, openging angle distribution,
visible energy distribution and so on.
As for the detection facility, Hyper-Kamiokande has been proposed with 
the fiducial volume 25 time larger than SK~\cite{HyperK}, and hence the number of 
events increases by the same amount, which improve the bound, roughly speaking, 
by a factor of five.
If the new customized analysis upgrades the sensitivity for $|\Theta_\mu |^2$ by 
a factor of two for instance, Hyper-Kamiokande will improve the bound by one order of 
magnitude and will reach the present accelerator bound. 
With the refinement of the analysis and the upgrade of the facility, 
there will be a good chance to discover the atmospheric sterile neutrino.

\subsection*{Acknowledgments}
We would like to thank Particle and Astroparticle Division of
Max-Planck-Institut f\"ur Kernphysik at Heidelberg for hospitality.
The works of T.A. and A.W. are supported by the Young Researcher Overseas Visits Program 
for Vitalizing Brain Circulation Japanese in JSPS (No.~R2209).
T.A. is also supported by KAKENHI (No. 21540260) in JSPS.
\bigskip

\appendix
\section{Details for the decay processes}
For the sake of completeness, we present here necessary formulas for the
decay processes to produce the results in Section~\ref{flux} and~\ref{events}.
In this Appendix, we flexibly use the symbol $\gamma$ and $x$ to denote the gamma 
factor of the parent particles and (twice of) the daughter energy in unit of the 
parent mass. The electron mass is neglected in the following formulas.
\subsection{$\mu^-  \to e^- \bar{\nu}_e N$}
\label{A1}
This decay process is conducted by the charged-current. In the laboratory frame,
the decay width $\Gamma'$ is
\begin{eqnarray}
\Gamma' \,=\,\frac{G_F^2 m_\mu^5 |\Theta_\mu|^2}{192\pi^3}I_0\,
\left(\frac{1}{\gamma} \right),
\end{eqnarray}
where
\begin{eqnarray}
I_0\,=\,\Big[\, 1-8r^2+8r^6-r^8-24r^4\ln(r)\,\, \Big], \quad
r = \frac{M_N}{m_\mu},\quad \gamma = \frac{E_\mu}{m_\mu}.
\end{eqnarray}
Here $E_\mu$ is the muon energy in the laboratory frame.
The energy distribution of the sterile neutrino $N$ is given by
(with $x = 2E_N/m_\mu$)
\begin{eqnarray}
\frac{1}{\Gamma'}\frac{d\Gamma'}{dx} = g(\gamma,x)=
  \left\{ 
    \begin{array}{l l}
      g_{\rm low}(\gamma, x) & (1 < \gamma < \gamma_{\rm cr},~ x^- < x < x^c )
      \\
      g_{\rm high}(\gamma, x) & (1 < \gamma < \gamma_{\rm cr},~  x^c < x < x^+)
      \\
      g_{\rm high}(\gamma, x) & (\gamma_{\rm cr} < \gamma ,~  x^- < x < x^+)
      \\ 
      0 & (\mbox{all others})
    \end{array}
  \right. \,,
\end{eqnarray}
where
\begin{eqnarray}
&&x^- = 2 \, r \,,~~~
  x^c = \gamma ( 1 + r^2) - \gamma \beta (1 - r^2 ) \,,~~~
  x^+ = \gamma ( 1 + r^2) + \gamma \beta (1 - r^2 ) \,,
  \nonumber \\
&& \gamma_{\rm cr} = \frac{1 + r^2}{2 \, r} \,.
\end{eqnarray}
The functions $g_{\rm low}$ and $g_{\rm high}$ are given by
\begin{eqnarray}
&&  g_{\rm low} (\gamma, x)
  \,=\, \frac{2 \sqrt{x^2 - 4 r^2}}{3 I_0}
  \Bigl[ \,
  (2-8 \gamma^2) x^2 + 9 (1+r^2) \gamma x
  + 4 r^2 (2 \gamma^2 - 5 ) \,
  \Bigr] \,,
  \nonumber \\
&&  g_{\rm high} (\gamma, x) \,=\, 
\frac{1}{6I_0 \sqrt{\gamma^2-1}}\Big[
(1+r^2)(5 + (36\gamma^2 -50)r^2 + 5r^4)
+24r^2 \gamma \left(\, 3 -2\gamma^2 \right)x \nonumber\\
&&\quad\quad\quad\quad\quad\quad-9 \left(2\gamma^2 -1 \right)(1+r^2)x^2
+ 4\gamma\left( 4\gamma^2 -3 \right)x^3 \,\Big]+\frac{1}{2}\,g_{\rm low}(\gamma, x).
\end{eqnarray}

\subsection{$\pi^+ \to \mu^+N$}
\label{A2}
The decay width in the laboratory frame is 
\begin{eqnarray}
\Gamma_\pi' \,=\, \frac{G_F^2 f_\pi^2 m_{\pi^\pm}^3 |V_{ud}|^2 |\Theta_\mu |^2 }{8\pi}
\beta_f \Big[\, r_N^2 + r_\mu^2 - (r_N^2 - r_\mu^2)^2 \,\Big]
\left(\frac{1}{\gamma}\right),
\end{eqnarray}
where
\begin{eqnarray}
\beta_f = \sqrt{1 - 2(r_N^2 + r_\mu^2) + (r_N^2 - r_\mu^2)^2},
\quad r_N = \frac{M_N}{m_{\pi^\pm}},\quad r_\mu = \frac{m_\mu}{m_{\pi^\pm}},\quad
\gamma = \frac{E_\pi}{m_{\pi^\pm}}.
\end{eqnarray}
The energy distribution of the sterile neutrino is given by
(with $x = 2E_N/m_{\pi^\pm}$)
\begin{eqnarray}
\frac{1}{\Gamma_\pi'}\frac{d\Gamma_\pi'}{dx} \,=\,
\left\{ \begin{array}{ll}
\frac{1}{2\beta_f}\frac{1}{\sqrt{\gamma^2 - 1}} & (1<\gamma,\,\, x^- < x < x^+) \\
0 & ({\rm all\,\, others})\\
\end{array}\right.,
\end{eqnarray}
where
\begin{eqnarray}
x^\pm \,=\,(1+r_N^2 - r_\mu^2)\gamma \pm \beta_f\sqrt{\gamma^2 - 1}.
\end{eqnarray}

\subsection{$N \to e^- e^+ \nu_\alpha$ ~($\alpha = \mu, \tau$)}
\label{A3}
This process is conducted by the neutral current if $\Theta_e $ contribution
is negligible. In this case, the decay width in the laboratory frame is given by
\begin{eqnarray}
\Gamma_N &\,=\,& 
\frac{G_F^2 |\Theta_\alpha |^2 M_N^5 }{192\pi^3}
\left( \frac{1}{4} - \sin^2\theta_W +2\sin^4\theta_W \right)
\left( \frac{1}{\gamma} \right).
\end{eqnarray}
In terms of the invariant mass $M_{ee}^2$, 
the visible energy $E_{ee}$ and the opening angle of $e^\pm$ momenta $\theta_{ee}$,
the differential decay width is written as 
\begin{eqnarray}
\frac{1}{\Gamma_N}\int\!\! d\Gamma_N = 
\int_0^1 \!dz_{ee}\,
\!\int_{x_v^-}^{x_v^+} 
\! dx_v  \int_{-1}^{1-\frac{8z_{ee}}{x_v^2}} 
\!\!\!\!d\!\cos\theta_{ee}\, J(z_{ee},x_v,\cos\theta_{ee})
\,K(\gamma,z_{ee},x_v,\cos\theta_{ee}),
\end{eqnarray}
where
\begin{eqnarray}
z_{ee} = \frac{M_{ee}^2}{M_N^2}, \quad
x_v = \frac{2E_{ee}}{M_N}, \quad
\gamma = \frac{E_N}{M_N},\quad
x_v^\pm =  \gamma(1+z_{ee})\pm \sqrt{\gamma^2 -1}(1-z_{ee}),
\end{eqnarray}
and
\begin{eqnarray}
J(z_{ee},x_v,\cos\theta_{ee}) \,=\,
 \frac{4z_{ee}}{\sqrt{x_v^2 - 4z_{ee}}\sqrt{x_v^2
(1-\cos\theta_{ee})^4- 8z_{ee}(1-\cos\theta_{ee})^3}},
\end{eqnarray}
\begin{eqnarray}
K(\gamma,z_{ee},x_v,\cos\theta_{ee}) =
\frac{3z_{ee}}
{\sqrt{\gamma^2-1}(1-\cos\theta_{ee})(x_v^2 - 4z_{ee})^2}\Big[
f_0 + f_1 \cos\theta_{ee} \Big].
\end{eqnarray}
Here $f_0$ and $f_1$ are 
\begin{eqnarray}
f_0 &=& 4z_{ee}\left( \, 3 + 2(10\gamma^2 - 1)z_{ee} - z_{ee}^2 \right)
-40\gamma z_{ee}(1+z_{ee})x_v  \nonumber\\
&&+4z_{ee}\left( (3-2\gamma^2) +2z_{ee} \right)x_v^2
+4\gamma(1+z_{ee})x_v^3 - (1+z_{ee})x_v^4,
\end{eqnarray}
\begin{eqnarray}
f_1 &=& 4z_{ee}\left( \, -1 + 2(2\gamma^2 - 1)z_{ee} + 3z_{ee}^2 \right)
-8\gamma z_{ee}(1+z_{ee})x_v  \nonumber\\
&&+4 \left(1+ (2\gamma^2+1)z_{ee} -z_{ee}^2 \right)x_v^2
-4\gamma(1+z_{ee})x_v^3 + (1+z_{ee})x_v^4.
\end{eqnarray}
By integrating over $\cos\theta_{ee}$, one finds for instance
\begin{eqnarray}
\frac{1}{\Gamma_N}\frac{d\Gamma_N}{dz_{ee} dx_v} &\,=\,&
\frac{1}{\sqrt{\gamma^2-1}}(1-z_{ee})(1+2z_{ee})\quad\quad 
(1<\gamma,\,\, 0 < z_{ee} <1,\,\, x_v^- <x_v <x_v^+),\nonumber\\
\frac{1}{\Gamma_N}\frac{d\Gamma}{dz_{ee}} &\,=\,&
2(1-z_{ee})^2(1+2z_{ee})\quad\quad\quad\quad\quad (0 < z_{ee} <1).
\label{Mee2Dis}
\end{eqnarray}
It is now straightforward to obtain~(\ref{MeeDis}) from~(\ref{Mee2Dis}).
The analytical expressions for $\theta_{ee}$ distribution are rather lengthy
and we do not present them here.



\begin{thebibliography}{99}

\bibitem{Seesaw}
P.~Minkowski,
Phys.\ Lett.\ B {\bf 67} (1977) 421;
T.~Yanagida,
in {\em Proc. of the Workshop on the Unified Theory
and the Baryon Number in the Universe}, 
Tsukuba, Japan, Feb.~13-14, 1979, p.~95, 
eds. O.~Sawada and S.~Sugamoto, 
(KEK Report KEK-79-18, 1979, Tsukuba); 
Progr.\ Theor.\ Phys.\ {\bf 64} (1980) 1103 ; 
M.~Gell-Mann, P.~Ramond and R.~Slansky, 
in {\em Supergravity}, 
eds. P.~van~Niewenhuizen and D.~Z.~Freedman
(North Holland, Amsterdam 1980);
P.~Ramond, 
in {\em Talk given at the Sanibel Symposium}, 
Palm Coast, Fla., Feb.~25-Mar.~2, 1979, preprint CALT-68-709
(retroprinted as hep-ph/9809459);
S.~L.~Glashow,
in {\em Proc. of the Carg\'ese  Summer Institute on Quarks and Leptons},
Carg\'ese, July 9-29, 1979, 
eds. M.~L\'evy et. al, , (Plenum, 1980, New York), p707;
  R.~N.~Mohapatra and G.~Senjanovic,
  Phys.\ Rev.\ Lett.\  {\bf 44} (1980) 912.

\bibitem{leptogenesis}
  M.~Fukugita and T.~Yanagida,
  Phys.\ Lett.\  B {\bf 174} (1986) 45.


\bibitem{DM}
  S.~Dodelson, L.~M.~Widrow,
  Phys.\ Rev.\ Lett.\  {\bf 72} (1994)  17-20;
  X.~-D.~Shi, G.~M.~Fuller,
  Phys.\ Rev.\ Lett.\  {\bf 82} (1999)  2832-2835;
  A.~D.~Dolgov, S.~H.~Hansen,
  Astropart.\ Phys.\  {\bf 16} (2002)  339-344;
  K.~Abazajian, G.~M.~Fuller, M.~Patel,
  Phys.\ Rev.\  {\bf D64} (2001)  023501;
  K.~Abazajian, G.~M.~Fuller, W.~H.~Tucker,
  Astrophys.\ J.\  {\bf 562 } (2001)  593-604;
  T.~Asaka, M.~Laine and M.~Shaposhnikov,
  JHEP {\bf 0606} (2006) 053;
  JHEP {\bf 0701} (2007)  091;
  M.~Laine, M.~Shaposhnikov,
  JCAP {\bf 0806} (2008)  031.

\bibitem{Asaka:2005an}
  T.~Asaka, S.~Blanchet, M.~Shaposhnikov,
  Phys.\ Lett.\  {\bf B631} (2005)  151-156.


\bibitem{palsar}
  A.~Kusenko, G.~Segre,
  Phys.\ Lett.\  {\bf B396} (1997)  197-200;
  Phys.\ Rev.\  {\bf D59} (1999)  061302;
  G.~M.~Fuller, A.~Kusenko, I.~Mocioiu, S.~Pascoli,
  Phys.\ Rev.\  {\bf D68} (2003)  103002;
  M.~Barkovich, J.~C.~D'Olivo, R.~Montemayor,
  Phys.\ Rev.\  {\bf D70} (2004)  043005;
  A.~Kusenko,
  Int.\ J.\ Mod.\ Phys.\  {\bf D13} (2004)  2065-2084;
  A.~Kusenko, B.~P.~Mandal, A.~Mukherjee,
  Phys.\ Rev.\  {\bf D77} (2008)  123009.


\bibitem{BAU}
  E.~K.~Akhmedov, V.~A.~Rubakov, A.~Y.~.Smirnov,
  Phys.\ Rev.\ Lett.\  {\bf 81} (1998)  1359-1362.


\bibitem{Asaka:2005pn}
  T.~Asaka, M.~Shaposhnikov,
  Phys.\ Lett.\  {\bf B620} (2005)  17-26.

\bibitem{BAU2}
  M.~Shaposhnikov,
  JHEP {\bf 0808} (2008) 008;
  T.~Asaka, H.~Ishida,
  Phys.\ Lett.\  {\bf B692} (2010)  105-113;
  L.~Canetti and M.~Shaposhnikov,
  JCAP {\bf 1009}, 001 (2010);
  T.~Asaka, S.~Eijima and H.~Ishida,
  arXiv:1112.5565 [hep-ph].


\bibitem{Gorbunov:2007ak}
  D.~Gorbunov and M.~Shaposhnikov,
  JHEP {\bf 0710} (2007) 015.


\bibitem{pi}
  R.~E.~Shrock,
  Phys.\ Rev.\  {\bf D24} (1981)  1232;
  F.~P.~Calaprice, D.~F.~Schreiber, M.~B.~Schneider, M.~Green, R.~E.~Pollock,
  Phys.\ Lett.\  {\bf B106} (1981)  175-178;
  R.~C.~Minehart, K.~O.~H.~Ziock, R.~Marshall, W.~A.~Stephens, M.~Daum, B.~Jost, P.~R.~Kettle,
  Phys.\ Rev.\ Lett.\  {\bf 52} (1984)  804-807;
  M.~Daum, R.~Frosch, W.~Hajdas, M.~Janousch, P.~R.~Kettle, S.~Ritt, Z.~G.~Zhao,
  Phys.\ Lett.\  {\bf B361} (1995)  179-183;
  R.~Bilger {\it et al.} [Karmen Collaboration],
  Phys.\ Lett.\  {\bf B363} (1995)  41-45;
  P.~Astier {\it et al.} [NOMAD Collaboration],
  Phys.\ Lett.\  {\bf B527} (2002)  23-28.


\bibitem{pi2}
 R.~Abela, M.~Daum, G.~H.~Eaton, R.~Frosch, B.~Jost, P.~R.~Kettle
 and E.~Steiner,
 Phys.\ Lett.\ B {\bf 105} (1981) 263   [Erratum-ibid.\ B {\bf 106}
(1981) 513];
 M.~Daum, B.~Jost, R.~M.~Marshall, R.~C.~Minehart, W.~A.~Stephens
 and K.~O.~H.~Ziock,
 Phys.\ Rev.\ D {\bf 36} (1987) 2624;
 D.~A.~Bryman and T.~Numao,
 Phys.\ Rev.\ D {\bf 53} (1996) 558.


\bibitem{pi3}
 M.~Daum, M.~Janousch, P.~R.~Kettle, J.~Koglin, D.~Pocanic, J.~Schottmuller, 
C.~Wigger, Z.~G.~Zhao,
  Phys.\ Rev.\ Lett.\  {\bf 85} (2000)  1815-1818.


\bibitem{K}
  Y.~Asano, R.~S.~Hayano, E.~Kikutani, S.~Kurokawa, T.~Miyachi, M.~Miyajima, Y.~Nagashima, T.~Shinkawa {\it et al.},
  Phys.\ Lett.\  {\bf B104} (1981)  84;
  R.~S.~Hayano, T.~Taniguchi, T.~Yamanaka, T.~Tanimori, R.~Enomoto, A.~Ishibashi, T.~Ishikawa, S.~Sato {\it et al.},
  Phys.\ Rev.\ Lett.\  {\bf 49} (1982)  1305.

\bibitem{K2}
T.~Yamazaki, in Proc. Neutrino'84 (Dortmund, 1984).
  
\bibitem{acc}
  F.~Bergsma {\it et al.} [CHARM Collaboration],
  Phys.\ Lett.\  {\bf B128} (1983)  361;
  G.~Bernardi, G.~Carugno, J.~Chauveau, F.~Dicarlo, M.~Dris, J.~Dumarchez, M.~Ferro-Luzzi, J.~M.~Levy {\it et al.},
  Phys.\ Lett.\  {\bf B166} (1986)  479;
  S.~A.~Baranov, Y.~.A.~Batusov, A.~A.~Borisov, S.~A.~Bunyatov, V.~Y.~.Valuev, A.~S.~Vovenko, V.~N.~Goryachev, M.~M.~Kirsanov {\it et al.},
  Phys.\ Lett.\  {\bf B302p} (1993)  336-340;
  A.~Vaitaitis {\it et al.} [NuTeV and E815 Collaborations],
  Phys.\ Rev.\ Lett.\  {\bf 83} (1999)  4943-4946.

\bibitem{acc2}
  G.~Bernardi, G.~Carugno, J.~Chauveau, F.~Dicarlo, M.~Dris, J.~Dumarchez, 
  M.~Ferro-Luzzi, J.~-M.~Levy {\it et al.},
  Phys.\ Lett.\  {\bf B203} (1988) 332.

  
\bibitem{Kusenko}
  A.~Kusenko, S.~Pascoli, D.~Semikoz,
  JHEP {\bf 0511} (2005) 028.

\bibitem{ATMSKI}
  Y.~Ashie {\it et al.} [Super-Kamiokande Collaboration],
  Phys.\ Rev.\  {\bf D71} (2005)  112005.


\bibitem{Lipari}
  P.~Lipari,
  Astropart.\ Phys.\  {\bf 1} (1993) 195-227;
  P.~Gondolo, G.~Ingelman and M.~Thunman,
  Astropart.\ Phys.\  {\bf 5} (1996) 309.


\bibitem{Honda}
  M.~Honda, T.~Kajita, K.~Kasahara, S.~Midorikawa,
  Phys.\ Rev.\  {\bf D52} (1995) 4985-5005.

\bibitem{ATMfluxes}
  G.~Battistoni, A.~Ferrari, P.~Lipari, T.~Montaruli, P.~R.~Sala, T.~Rancati,
  Astropart.\ Phys.\  {\bf 12} (2000)  315-333;
  M.~Honda, T.~Kajita, K.~Kasahara, S.~Midorikawa,
  Phys.\ Rev.\  {\bf D70} (2004) 043008;
  M.~Honda, T.~Kajita, K.~Kasahara, S.~Midorikawa, T.~Sanuki,
  Phys.\ Rev.\  {\bf D75} (2007)  043006;
  M.~Honda, T.~Kajita, K.~Kasahara, S.~Midorikawa,
  Phys.\ Rev.\  {\bf D83} (2011)  123001.

\bibitem{BBN}
  A.~D.~Dolgov, S.~H.~Hansen, G.~Raffelt and D.~V.~Semikoz,
  Nucl.\ Phys.\ B {\bf 590} (2000) 562.

\bibitem{Nsearch1}
  A.~Atre, T.~Han, S.~Pascoli and B.~Zhang,
  JHEP {\bf 0905} (2009) 030;
  T.~Asaka, S.~Eijima and H.~Ishida,
  JHEP {\bf 1104} (2011) 011.

\bibitem{Ruchayskiy:2011aa}
  O.~Ruchayskiy and A.~Ivashko,
  arXiv:1112.3319 [hep-ph].


\bibitem{Ishihara}
See for example,
  C.~Ishihara,
  PhD Thesis, University of Tokyo, Feb. 2010,
  ``Full three flavor oscillation analysis of atmospheric neutrino data observed in Super-Kamiokande''.

\bibitem{HyperK}
  K.~Abe, T.~Abe, H.~Aihara, Y.~Fukuda, Y.~Hayato, K.~Huang, A.~K.~Ichikawa and M.~Ikeda {\it et al.},
  arXiv:1109.3262 [hep-ex].


\end{thebibliography}
\end{document}